\documentclass[]{article}

\usepackage{graphicx}
\usepackage{cancel}
\usepackage{authblk}
\usepackage{amsmath}
\usepackage{amsthm}
\usepackage{amssymb}

\usepackage[sort&compress,numbers]{natbib}
\usepackage[colorlinks=true,linkcolor=black, citecolor=blue, urlcolor=blue]{hyperref}

\begin{document}

\title{Potential-Based Formalism for Electrodynamics of Media with Weak Spatial Dispersion} 

\author[1,2]{Y. Solyaev}

\affil[1]{Institute of Applied Mechanics of Russian Academy of Sciences, Moscow, Russia}
\affil[2]{Moscow Aviation Institute, Moscow, Russia}

\setcounter{Maxaffil}{0}
\renewcommand\Affilfont{\itshape\small}

\date{\today}

\maketitle

\begin{abstract}
In this work, we develop a potential-based formalism for Maxwell's equations in isotropic media with weak spatial dispersion within the electric quadrupole-magnetic dipole approximation.
We introduce an operator form of the constitutive relations along with a modified Lorenz gauge condition, which enables the derivation of decoupled generalized wave equations for electromagnetic potentials.
For time-harmonic processes, we derive the representation of general solution for these equations as a combination of solutions to Helmholtz-type equations, whose parameters are determined by both standard and hyper-susceptibilities of the medium. 
We show that the proposed approach can be extended to more general constitutive relations and it provides a convenient framework for solving various applied problems.
Specifically, using a derived closed-form solution for the problem of plane wave incidence on a planar interface, we demonstrate that a correct definition of the Poynting vector within the multipole theory must incorporate quadrupole effects -- an aspect overlooked in some previous works that has led to inconsistent results.
We further establish the necessity of accounting for both propagated and evanescent longitudinal components in reflected and transmitted waves. The presence of these components, which follow directly from the general solution for electromagnetic potentials, is essential for satisfying all classical and additional boundary conditions in media with quadrupolar response (e.g., in metamaterials or quadrupolar liquid mixtures).
The complete set of these boundary conditions is derived based on the least action principle, ensuring variational consistency with the field equations and generalizing previously known formulations of multipole theory.
\end{abstract}

%%%%%%%%%%%%%%%%%%%%%%%%%%%%%%%%%%%%%%%%%%%%%%
\section{Introduction}

The electromagnetic theory with multipoles is a natural generalization of classical Maxwell equations in continuum media \cite{raab2005multipole,graham1992multipole}. The formulation of multipole theory implies that the energy density of a medium depends not only on the electric and magnetic fields but also on their spatial gradients. These gradients contribute to the constitutive equations and lead to modifications of Gauss's law and Amp\'ere's law \cite{raab2005multipole, slavchov2014quadrupole}. Such generalized continuum models are also known as the models of media with weak spatial dispersion \cite{simovski2018composite}. 

The framework of multipole theory enables the description of specific phenomena observed in anisotropic media such that natural optical activity \cite{landau2013electrodynamics,raab2005multipole}, gyrotropic  birefringence \cite{hornreich1968theory, graham1992magnetic} and Lorenz birefringence \cite{graham1990light}, etc. The models with weak spatial dispersion are widely involved in the description of metamaterials and metasurfaces  \cite{simovski2018composite, achouri2023spatial, petschulat2008multipole, dong2013all, yaghjian2013homogenization, chipouline2017analytical,liu2021effective}.
Within the quasi-static approximation, the multipole theory was used in the problems of chemical physics related to the refined analysis of solvation and self-salting-out of electrolytes, assessments on the Born energy, partial molar volume, and partial molar entropy of dissolved ions \cite{chitanvis1996continuum, jeon2003continuum, slavchov2014quadrupole, slavchov2014quadrupole2, dimitrova2020quadrupolarizability}, etc. Furthermore, in a coupled electromechanical formulation under quasi-static assumptions, the theory with quadrupolar polarization finds extensive applications in the modelling of piezoelectric materials and composites \cite{yang2006review, maugin2013continuum, yue2014microscale, solyaev2021electric, kalpakides2002material}.

The modern formulation of the multipole theory of various order was established in Ref. \cite{graham1990light} (see also \cite{raab2005multipole}). 
Subsequent discussion regarding the form and methods of deriving the additional boundary conditions (required due to the higher order of the field equations) was carried out in Refs. \cite{golubkov1995boundary, graham2000multipole, de2013electromagnetic, silveirinha2014boundary, yaghjian2016additional}. A well-posedness of the problem for electromagnetic media with quadrupolar response has been proved in Ref. \cite{bosello2003well} considering thermodynamic restrictions on the material constants. 
The problem of the formulation of constitutive relations that obey the material objectivity condition in multipole theory were the subject of intensive research (see \cite{raab2005multipole} and references therein). 
Corresponding discussion and reformulation of origin-independent constitutive equations of multipole theory was given in Refs. \cite{lange2003completion, de2006origin, achouri2023spatial, de2010reply, anelli2015origin}. A generalization of the reciprocity theorem and Poynting's theorem for a theory incorporating the quadrupole effects in electric polarization and magnetization was recently considered \cite{achouri2021extension}. We can also refer to the work by Kafadar \cite{kafadar1971theory}, where the field equations for the multipole theory of arbitrary order were derived based on the least action principle taking into account the relativistic effects. 

The present paper addresses the problem of defining electromagnetic potentials and deriving the corresponding wave equations within the framework of multipole theory. To the best of our knowledge, this problem has not been addressed previously. While the introduction of a scalar potential for the electric field in quadrupolar theory has been explored, it has been confined to the electrostatic approximation \cite{kafadar1971theory,chitanvis1996continuum, slavchov2014quadrupole, slavchov2014quadrupole2, dimitrova2020quadrupolarizability, slavchov2015polarized,solyaev2021electric}. It was shown that the corresponding generalized Gauss law in terms of scalar potential becomes a composition of two operators -- classical Laplacian and modified Helmhotz operator, which parameter ("quadrupolar length" \cite{dimitrova2020quadrupolarizability}) is defined via the ratio between standard and hyper-permittivity of dielectric media. 

In the present work, we introduce an operator formulation of the constitutive equations along with a modified Lorenz gauge condition to obtain uncoupled generalized wave equations for the electric scalar and vector potentials, thus extending potential-based methods to multipole theory.
The obtained generalized wave equations allow for a simple representation of the general solution in the case of time-harmonic processes. This solution takes the form of a combination of general solutions to the Helmholtz and modified Helmholtz equations, whose representations are well-known in Cartesian and various curvilinear coordinate systems. The particular solutions for the electromagnetic fields are defined via the derived generalized form of dyadic Green's functions. As a result, we provide the convenient method that allows to find analytical solutions for a wide class of applied problems within the electrodynamic theory with multipoles. 

In this work, we consider an example of isotropic media within the quadruple electric-dipole magnetic approximation. This media can be, e.g. the metamaterial consisted of randomly oriented particles  distributed in the isotropic matrix \cite{simovski2018composite} or the quadrupolar liquid mixture \cite{dimitrova2020quadrupolarizability}. It will be also shown, that  presented operator approach for the construction of general solution can be also extended for the more general class of materials, which constitutive equations can be presented in operator form (including, e.g.   anisotropic and chiral materials). We can also note that similar operator approaches for the theories with the high-order field equations were used previously within the generalized electrodynamics of free space \cite{lazar2014gradient,lazar2020second, lazar2020second2} as well as within the second gradient elasticity \cite{lurie2011eshelby, solyaev2023semi, solyaev2024complete, rezaei2024procedure, dell2024deformation}.

The example of application of the considered approach is shown within a rigorous derivation of analytical solution for the problem of reflection and transmission of plane wave at the planar boundary between two non-magnetic dielectrics. This problem was solved previously for isotropic \cite{yaghjian2016additional, yaghjian2018power, silveirinha2014boundary, simovski2018composite} and anisotropic (see \cite{raab2005multipole} and references therein) materials with multipole effects, although in the derivation of these solutions some parts of the reflected and transmitted waves were always neglected. Namely, as revealed from the derived form of general solution, the reflected and transmitted waves will contain two additional longitudinal components. One of these components is the propagated wave and another one is an evanescent wave localized around the boundary (in the transition layer \cite{silveirinha2014boundary}). 
The existence of such nonuniform waves in media with weak spatial dispersion is well-known \cite{agranovich1966spatial, landau2013electrodynamics,silveirinha2014boundary}. However, rigorous analytical solutions that fully account for these wave components have not been previously derived. Obtaining such a complete solution requires an extended set of boundary conditions, including an additional continuity condition for the quadrupole tensor \cite{graham2000multipole,silveirinha2014boundary} as well as conditions for the scalar potential and its normal gradient.
Previously, the need for such an extended number of boundary conditions was not discussed within the electrodynamic theory with multipoles. Although, the conditions for the scalar potential and its normal gradient in the case of electrostatics with quadrupoles were taken into account, e.g. in Refs. \cite{slavchov2015polarized, yue2014microscale, solyaev2021electric}. 

In the present study, we show that for oblique incidence problem, the theory necessitates six continuity conditions. The form of these boundary conditions is obtained together with the field equations based on the least action principle. We use a non-relativistic approximation and show that the consistent form of boundary conditions (that were widely discussed in multipole theory \cite{raab2005multipole, graham2000multipole, de2013electromagnetic, silveirinha2014boundary, yaghjian2016additional}) are directly follows from variational approach. Note that previously, the form of solutions for the oblique incidence problems were obtained by neglecting from one to four additional continuity conditions \cite{raab2005multipole, yaghjian2016additional, yaghjian2018power, silveirinha2014boundary} that results in the approximate nature of these solutions. 
Even in the case where one of the media is a vacuum, the oblique incidence problem gives rise to one purely transverse reflected wave in the vacuum and three waves in the quadrupole material: a propagating transverse wave, a propagating longitudinal wave, and an evanescent longitudinal wave. Thus, even in this simplified scenario, we require four boundary conditions, which has never been accounted for before within multipole theory \cite{raab2005multipole, yaghjian2016additional, yaghjian2018power, silveirinha2014boundary}.
For the derived new complete solution we show that the longitudinal waves do not take large amount of energy in the total energy balance, though they always exist and should be considered as an important property of spatially dispersive media \cite{agranovich1966spatial,landau2013electrodynamics}. 

Additional aspect that is discussed in the present study, is the definition of Poynting vector within the multipole theory. Based on the obtained closed-form analytical solution for the normal incidence problem, we show that the definition of Poynting vector should be given accounting for the contribution of quadrupole effects. 
This definition was derived within the considered theory for the first time in Refs. \cite{yaghjian2016additional, yaghjian2018power} through generalized time and spatial averaging of the energy flux, following standard procedures for media with spatial dispersion \cite{landau2013electrodynamics, silveirinha2009poynting}. However, several works have neglected the quadrupolar terms in the Poynting vector definition. For instance, Ref. \cite{de2006surprises} omitted these terms when analyzing position-dependent artifacts in energy transfer in the media exhibiting multipole response -- an omission that may explain the reported artifacts. More recently, discussion of the Poynting theorem in multipole electrodynamics \cite{achouri2021extension} have also employed the classical Poynting vector definition without quadrupole contributions, though this simplification did not affect the final form of the balance relations in that particular context.
In the present study, we show that without the contribution of quadrupole effects on the Poynting vector, the energy balance in the reflection/transmission problems cannot been preserved and the errors become significant for the high frequency ranges. Note that previously, this balance was evaluated based on numerical calculations \cite{yaghjian2016additional, yaghjian2018power, silveirinha2014boundary}. In the present study, the corresponding explicit analytical solution is given.
Also, we show that the correct definition of Poynting vector can be simply obtained based on the Poynting theorem that directly follows from the field equations of multipole theory (see Appendix B). 

The rest part of the paper is organized as follows. In Section 2 we derive the field equations and corresponding boundary conditions of the theory based on the least action principle. The operator form of constitutive equations is  introduced and discussed in this section. In Section 3 the uncoupled generalized wave equations and their general solution for the electromagnetic potentials are derived. In Section 4 we consider the propagation phenomena and discuss the correct definitions for the Poynting vector within the considered theory. In Section 5 we provide the complete analytical solution for the reflection/transmission problem and show the peculiarities of this solution in the multipole theory. 
\newpage

%%%%%%%%%%%%%%%%%%%%%%%%%%%%%%%%%%%%%%%%%%%%%%
\section{Field equations and boundary conditions}

We consider a non-absorbing rigid medium exhibiting dipolar and quadrupolar electric polarization and dipolar magnetization. From the viewpoint of multipole expansion, the contributions of electric quadrupoles and magnetic dipoles are of the same order \cite{raab2005multipole, graham1992magnetic}. Thus, the considered variant of the theory is the most simple generalization of classical Maxwell equations in continuum media within the class of multipole theories. The coordinate system is the rest frame and the relativistic effects are out of consideration. 
The Lagrangian density of the media can be defined as follows:
\begin{equation}
\label{L}
	\mathcal L= \mathcal L_{f} + \mathcal L_{i} + \mathcal L_{m}
\end{equation}
where the Lagrangian density of the electromagnetic field is given by
\begin{equation}
\label{Lf}
	\mathcal L_{f}= 
	\tfrac{\varepsilon_0}{2} \textbf E^2
	- \tfrac{1}{2\mu_0} \textbf B^2
\end{equation}
in which the electric and magnetic fields are defined by the scalar potential $\phi$ and the vector potential $\textbf A$ as usual:
\begin{equation}
\label{pot}
\begin{aligned}
	\textbf E= -\dot{\textbf A}-\nabla\phi,\qquad
	\textbf B= \nabla\times\textbf A
\end{aligned}
\end{equation}
where the dot defines the time derivative $\dot{\textbf A} = \partial\textbf A /\partial t$.

The Lagrangian density related to the interaction between the field and free charges $\rho$ and currents $\textbf J$ is
\begin{equation}
\label{Li}
	\mathcal L_{i}= -\phi\rho + \textbf A\cdot \textbf J
\end{equation}

The  Lagrangian density related to the polarization and magnetization of material is defined accounting for the contribution of quadrupoles:
\begin{equation}
\label{Lm}
	\mathcal L_{m}= 
	\textbf P\cdot\textbf E 
	+ \textbf M\cdot\textbf B
	+ \textbf Q:\nabla\textbf E
\end{equation}
where $\textbf P$, $\textbf M$ are the vectors of electric polarization and magnetization, respectively, and $\textbf Q$ is the second rank tensor of electric quadrupolar polarization that is work conjugate to the gradient of electric field (see, e.g. \cite{slavchov2014quadrupole,achouri2021extension}). The double dot product symbol denotes the full contraction of second rank tensors, i.e. $\textbf Q:\nabla\textbf E = \sum_{i,j=1}^3Q_{ij}E_{i,j}$.

Introduction of the electromagnetic potentials \eqref{pot} provides the fulfilment of  two homogeneous Maxwell equations:
\begin{equation}
\label{Maxh}
\begin{aligned}
	\nabla\times\textbf E= -\dot{\textbf B},\qquad
	\nabla\cdot\textbf B= 0
\end{aligned}
\end{equation}

The remaining two Maxwell equations together with corresponding boundary conditions can be obtained based on the least action principle similarly to classical approach (see, e.g. \cite{landau2013electrodynamics,loudon1997propagation,auffray2015analytical}). Substituting \eqref{pot} into \eqref{L}, \eqref{Lf}, \eqref{Li}, \eqref{Lm}, we obtain:
\begin{equation}
\label{potL}
\begin{aligned}
	\mathcal L&= 
	\tfrac{\varepsilon_0}{2}\left(\dot{\textbf A}+\nabla\phi\right)^2
	- \tfrac{1}{2\mu_0}\left(\nabla\times\textbf A\right)^2
	-\phi\rho + \textbf A\cdot \textbf J\\
	&-\textbf P\cdot\left(\dot{\textbf A}+\nabla\phi\right) 
	+ \textbf M\cdot(\nabla\times\textbf A)
	-\textbf Q:\left(\nabla\dot{\textbf A}+\nabla\nabla\phi\right)
\end{aligned}
\end{equation}

Then, the statement of the least action principle can be defined as follows:
\begin{equation}
\label{S}
\begin{aligned}
	&\delta \mathcal S =0, \qquad
	\mathcal S= 
	\int_{t_0}^{t_1}\int_\Omega \mathcal L dV dt,\\
	&\delta \mathcal S = \int_{t_0}^{t_1}\int_\Omega 
	\Big(
		-\varepsilon_0\textbf E\cdot\left(\delta\dot{\textbf A}+\nabla\delta\phi\right)
	- \tfrac{1}{\mu_0}\textbf B\cdot\left(\nabla\times\delta\textbf A\right)	
	-\rho\,\delta\phi + \textbf J\cdot\delta\textbf A\\
	&-\textbf P\cdot\left(\delta\dot{\textbf A}+\nabla\delta\phi\right) 
	+ \textbf M\cdot(\nabla\times\delta\textbf A)
	-\textbf Q:\left(\nabla\delta\dot{\textbf A}+\nabla\nabla\delta\phi\right)
	\Big)
	 dV dt
\end{aligned}
\end{equation}
where $S$ is action functional and the spatial integration is performed over the domain $\Omega$ with boundary $\partial\Omega$ that can contain sharp edges $\partial\partial\Omega$. The time integration is performed over time interval $[t_0,t_1]$ with initial conditions being prescribed at the time moment $t_0$.

Through the application of integral theorems and appropriate vector calculus identities, the variation of the action functional \eqref{S} can be expressed in the following form (detailed derivations are provided in Appendix A):
\begin{equation}
\label{Sder}
\begin{aligned}
	\delta \mathcal S &= \int_{t_0}^{t_1}\int_\Omega 
	\Big(
		(\nabla\cdot\textbf D-\rho)\,\delta\phi	
		+ (\dot{\textbf D}-\nabla\times\textbf H + \textbf J)\cdot\delta\textbf A
	\Big)
	 dV dt \\
	 &+ \int_{t_0}^{t_1}\int_{\partial\Omega} 
	\Big(
		-(\textbf n\cdot\textbf D-\nabla_S\cdot(\textbf n\cdot\textbf Q) - K Q_{nn})\,\delta\phi\\
		&\qquad\qquad\quad\,\,\,-Q_{nn}\,\delta(\partial_n\phi)
		+\textbf n\times \left(\textbf H+(\textbf n\cdot\dot{\textbf Q})\times\textbf n\right)\cdot\delta\textbf A
	\Big)dS dt\\
	&- \int_{t_0}^{t_1}\int_{\partial\partial\Omega} 
	 [\textbf n\cdot\textbf Q\cdot \pmb \nu]\, \delta\phi \,dL dt
\end{aligned}
\end{equation}
where $\textbf D = \varepsilon_0\textbf E+\textbf P
		- \nabla\cdot\textbf Q$ is the total electric displacement vector that takes into account the dipole and quadrupole polarization of the media \cite{raab2005multipole}; 
		$\nabla_S = \nabla - (\textbf n\cdot\nabla)\textbf n$ is the surface gradient operator; 
		$Q_{nn}=\textbf n\cdot\textbf Q\cdot \textbf n$ is the normal component of quadrupole tensor on the body boundary;
		$K = - \nabla\cdot\textbf n$ is twice the mean curvature of the boundary $\partial \Omega$;
		$\partial_n \phi = (\textbf n\cdot\nabla)\phi$ is the normal gradient of scalar potential on the body boundary;
		$\textbf n$ is the outward unit normal to the body boundary $\partial\Omega$; 
		the brackets $[...]$ denote the jump of the enclosed quantities across the edge; 
		$\pmb \nu$ is the co-normal vector that is tangent to surface $\partial\Omega$ and normal to edge $\partial\partial\Omega$.
        
Note that considering the infinite domain $\Omega$ one should avoid in \eqref{Sder}, \eqref{Ssl} the terms  that are integrated over its boundary $\partial\Omega$ and edges $\partial\partial\Omega$ \cite{landau2013electrodynamics}. However, for the finite size domain, these surface terms should be remained and they provide us the definition of the boundary conditions that will be consistent with the corresponding field equations from the viewpoint of variational approach \cite{eremeyev2018applications}.   The initial boundary conditions are neglected in \eqref{Sder} since we will consider only the time-harmonic processes in this study. The form of obtained generalized initial conditions for multipole theory is presented in Appendix A.

Taking into account the structure of surface and line integrals in \eqref{Sder}, we can add the appropriate part of action functional related to the work of electromagnetic field over the surface charge $\bar{\rho}$, surface current $\bar{\mathbf{J}}$, surface quadrupole density $\bar{Q}$, and over the charges distributed along the body edges $q_e$. This additional part should be added in \eqref{Sder} and it has the form:
\begin{equation}
\label{Ssl}
\begin{aligned}
	\delta\mathcal S_e &= \int_{t_0}^{t_1}\int_{\partial\Omega} 
	\Big(
		\bar \rho\delta\phi +\bar Q\delta(\partial_n\phi)
		-(\textbf n\times \bar {\textbf J})\cdot\delta\textbf A
	\Big)dS dt
	+ \int_{t_0}^{t_1}\int_{\partial\partial\Omega} 
	\rho_e \delta\phi \,dL dt
\end{aligned}
\end{equation}

Taking into account the independence of variations of scalar $\delta\phi$ and vector $\delta\textbf A$ potentials in the body volume and on its boundary, as well as the independence of normal derivative of scalar potential $\delta(\partial_n\phi)$ on the body boundary, from \eqref{Sder}, \eqref{Ssl} we obtain the following formulation of the boundary value problem of multipole theory under electric-quadrupole/magnetic dipole approximation.
The field equations have the same to classical form, although the constitutive equations for $\textbf D$ field are more general within the considered theory:
\begin{subequations}
\label{fe}
\begin{align}
	\textbf x\in\Omega:\quad
	&\nabla\times\textbf E= -\dot{\textbf B}, \label{fea}\\
	&\nabla\cdot\textbf B= 0, \label{feb}\\
	&\nabla\cdot\textbf D = \rho, \label{fec}\\
	&\nabla\times\textbf H = \textbf J + \dot{\textbf D} \label{fed}
\end{align}
\end{subequations}
where two first equations are the homogeneous Maxwell equations \eqref{Maxh} that remain unchanged within the multipole theory \cite{raab2005multipole,kafadar1971theory}; the last two equations equations are the generalized Gauss law and Amp\'ere law that follows from \eqref{Sder} as the Euler-Lagrange equations defined on the variations of scalar and vector potentials in the body volume, respectively. 
 
 The full set of boundary conditions for the multipole theory is given by:
\begin{subequations}
\label{bcs}
\begin{align}
	\textbf x\in\partial\Omega:\quad
	&\textbf n\cdot\textbf D-\nabla_S\cdot(\textbf n\cdot\textbf Q) 
	- K Q_{nn}=\bar \rho 
	\quad or \quad \phi = \bar \phi \label{bcsa}\\
	& \textbf n\times\textbf H
	+\textbf n\cdot\dot{\textbf Q}\cdot(\textbf I - \textbf n\otimes\textbf n)=\bar {\textbf J}
	\quad or \quad \textbf n\times\textbf A = \bar {\textbf A} \label{bcsb}\\
		&Q_{nn} = \bar Q\quad or \quad \partial_n\phi =\bar E \label{bcsc}\\
	&\textbf n\cdot\textbf B = 0, \quad \textbf n\times\textbf E = 0 \label{bcsd}
\end{align}
\end{subequations}
where we separate by "$or$" the natural and essential boundary conditions that follow from the surface integral terms in \eqref{Sder}, \eqref{Ssl}; also we add two  boundary conditions \eqref{bcsd} that do not follow from the least action principle, however that can be obtained following standard integration methods for the homogeneous Maxwell equations \eqref{fea}, \eqref{feb} \cite{silveirinha2014boundary,yaghjian2016additional}. Also, we note that the form of \eqref{bcsb} is obtained from \eqref{Sder} by using standard vector identities that are given in Appendix A.

The natural conditions in \eqref{bcsa}, \eqref{bcsb} are the generalization of classical boundary conditions for the electric displacement field and for magnetic field. The corresponding essential boundary conditions in \eqref{bcsa}, \eqref{bcsb} provides possibility to prescribe the scalar potential and the tangential components of vector potential on the body boundary. 
In the case of flat boundaries, the form of natural conditions in \eqref{bcsa}, \eqref{bcsb} coincides with those that were derived previously based on the integration of field equations of multipole theory \cite{raab2005multipole, silveirinha2014boundary, yaghjian2016additional}. However, for the curved boundaries, the least action principle provides us additional term related to the mean curvature of the boundary $K$ \eqref{bcsa}. This term will be avoided if the normal component of quadruple tensor $Q_{nn}$ is assumed to be zero on the free boundary. 

Additional boundary condition for the normal component of quadrupole tensor $Q_{nn}$  \eqref{bcsc} was derived within the multipole theory in Refs. \cite{graham2000multipole,silveirinha2009poynting,yaghjian2013homogenization} based on integration of field equations. In the present study, we show that this condition is the natural boundary condition that is defined on the variation of normal gradient of scalar potential $\delta(\partial_n\phi)$ on the body boundary (see \eqref{Sder}). Usually, it is assumed that this term should be equal to zero (i.e. $\bar Q = 0$) \cite{silveirinha2009poynting}.  The corresponding essential condition can be used to prescribe $\partial_n\phi$. In the case of electrostatics, this condition is reduced to the prescribed normal component of electric field $\bar E$ or its continuity over the contact boundary between two dielectrics \cite{slavchov2015polarized, yue2014microscale, solyaev2021electric}. In the  electrodynamics with multipoles this additional condition together with standard condition for the scalar potential were out of consideration in all previous studies, although they become important if we want to derive the complete solutions accounting for the non-uniform nature of reflected/transmitted waves. Such solution is presented below in Section 5.

Finally, the set of surface boundary conditions within the multipole theory should be accompanied by the boundary conditions on edges, that are given by:
\begin{equation}
\label{bce}
\begin{aligned}
	\textbf x\in\partial\partial\Omega:\quad
	&[\textbf n\cdot\textbf Q\cdot \pmb \nu] = q_e
	\quad or \quad \phi = \phi_e
\end{aligned}
\end{equation}
where we see that $q_e$ is formally the line density of quadrupoles (though it has the dimension and physical meaning of charge line density) and $\phi_e$ is the electric potential that can be prescribed directly on the sharp edge. These conditions follows from the line integral that arise in the variational formulation \eqref{Sder}, \eqref{Ssl}. 

For the bodies with smooth boundaries, the edge-type boundary conditions \eqref{bce} do not arise. Although, for the problems with non-smooth boundaries they become important. Similar examples can be found in high-grade elasticity theories for the problems with sharp wedges  \cite{solyaev2022elastic,reiher2017finite}.
Within the electrodynamics these conditions were not discussed previously (for the best of author's knowledge), though they can be important for the description of edge polarization and corner charges in artificial metamaterials \cite{ren2021quadrupole}.

Thus, for the flat and smooth boundaries the variational approach leads to a formulation of boundary value problem \eqref{fe}-\eqref{bce} that coincides with the previously known one. However, this approach provides clarifications for curved boundaries and for the domains with edges. Following least action principle, we derived a complete set of boundary conditions applicable to solving multipole theory problems \eqref{bcs}. It should be noted that this set of boundary conditions is not totally independent. Specifically, conditions \eqref{bcsb} and \eqref{bcsd} may be employed interchangeably based on the convenience in some particular boundary value problems.
It should be also noted that the presented form of boundary value problem for the theory with electric quadrupoles \eqref{fe}-\eqref{bce} is independent on constitutive equations and it will be also valid in the case of  anisotropic materials.

The following derivations require the continuity conditions at a flat boundary between two quadrupolar materials. From \eqref{bcs} we obtain the following continuity conditions (as mentioned above, not all of these conditions are independent):
\begin{equation}
\label{bcc}
\begin{aligned}
	&[\textbf n\cdot\textbf D-\nabla_S\cdot(\textbf n\cdot\textbf Q)]= 0, \quad
	[\phi] = 0\\
	& [\textbf n\times\textbf H
	+\textbf n\cdot\dot{\textbf Q}\cdot(\textbf I - \textbf n\otimes\textbf n)] = 0, \quad
	[\textbf n\times\textbf A] = 0\\
		& [Q_{nn}] = 0, \quad [\partial_n\phi] = 0,\quad
		[\textbf n\cdot\textbf B] = 0, \quad [\textbf n\times\textbf E] = 0
\end{aligned}
\end{equation}

We now define the constitutive equations for the isotropic theory. In a linear medium, the current density, polarization, and magnetization are given by the standard relations involving the conductivity $\sigma$ and the electric $\chi_e$ and magnetic $\chi_m$ susceptibilities:
\begin{equation}
\label{ce}
\begin{aligned}
	\textbf J= \sigma \textbf E, \qquad
	\textbf P=\varepsilon_0\chi_e\textbf E, \qquad
	\textbf M=\chi_m  \textbf H = \tfrac{\chi_m}{\mu}  \textbf B,\qquad\textbf B= \mu_0(\textbf H+\textbf M)
\end{aligned}
\end{equation}
where $\mu = (1+\chi_m)\mu_0$ is the magnetic permeability of media.

In the case of isotropy, the second rank tensor of quadrupoles density can be defined as follows:
\begin{equation}
\label{ceq}
\begin{aligned}
	\textbf Q& = \alpha_1 \textbf I \nabla\cdot\textbf E
				+ \alpha_2 (\nabla\textbf E + \textbf E\nabla)
\end{aligned}
\end{equation}
where $\alpha_1$, $\alpha_1$ are the material hyper-susceptibilities, $\textbf I$ is the identity (unit) second rank tensor. 

Usually, the tensor of quadrupoles is assumed to be traceless \cite{raab2005multipole,silveirinha2014boundary,slavchov2014quadrupole} that implies that we should use $\alpha_1 =-2\alpha_2/3$ in \eqref{ceq}. Although, there exist the variants of constitutive equations for metamaterials \cite{simovski2018composite,agranovich2006spatial} and for fluids \cite{jeon2003continuum} with non-zero trace ($\alpha_1\neq\alpha_2\neq0$) and for metamaterials with zero effects related to the electric field divergence ($\alpha_1=0$) \cite{simovski2018composite}. Thus, in the present study we consider the general two-parametric definition of the constitutive equations for quadrupole tensor $\textbf Q$, though it can be easily reduced to the appropriate particular case by using corresponding values of hyper-susceptibilities $\alpha_i$. Also, we assume that tensor $\textbf Q$ is symmetric since in the case of electrostatics the electric field is the potential field ($E_i = -\phi_{,i}$) and it is valid that $E_{i,j}=E_{j,i} = -\phi_{,ij}$. Therefore, the antisymmetric part of the electric field gradient was not included in \eqref{ceq}.

The constitutive equations for the electric displacement field (introduced in \eqref{Sder}) can be represented by using \eqref{ceq},  as follows:
\begin{equation}
\label{D}
\begin{aligned}
	\textbf D &= \varepsilon_0\textbf E+\textbf P
		-\nabla\cdot\textbf Q
		= \varepsilon\textbf E-(\alpha_1+\alpha_2) \nabla \nabla\cdot\textbf E - \alpha_2 \nabla^2\textbf E
		= \varepsilon L \textbf E
\end{aligned}
\end{equation}
where $\varepsilon = \varepsilon_0(1 + \chi_e)$ is the medium permittivity, and we introduce the vector differential operator $L$ that relates the electric field $\mathbf{E}$ to the electric displacement $\mathbf{D}$. This operator can be defined as follows:
\begin{equation}
\label{Lo}
\begin{aligned}
	L(...) 	
	&= 
	(...)
	-(l_1^2-l_2^2) \nabla \nabla\cdot (...)
	- l_2^2 \nabla^2 (...)\\
	&= 
	(...)
	- l_1^2 \nabla \nabla\cdot (...)
	+ l_2^2 \nabla\times\nabla\times (...)\\
	&= 
	(...)
	- l_1^2 \nabla^2 (...)
	- (l_1^2-l_2^2) \nabla\times\nabla\times (...)\\
\end{aligned}
\end{equation}
where we use standard vector calculus identities for vector Laplacian and introduce two length scale parameters $l_1 = \sqrt{(\alpha_1+2\alpha_2)/\varepsilon}$ and  $l_2 = \sqrt{\alpha_2/\varepsilon}$. 

Note that in the definition for $\textbf D$ field \eqref{D}, the customary factor of 1/2 preceding the quadrupole term (commonly employed in multipole theory \cite{raab2005multipole}) is absent. We omitted this factor to simplify the formulations of the least action principle. However, it can be readily recovered by means of appropriate renormalization of the material constants $\alpha_i$.

In the case of zero values of hyper-susceptibilities $\alpha_i=0$ (and $l_i=0$), the operator relation \eqref{D} reduces to classical linear constitutive equation for isotropic materials $\textbf D = \varepsilon \textbf E$. Non-zero length scale parameters corresponds to the weakly non-local effects in the media and results in the differential form of constitutive equations for $\textbf D$ field. Note that even in the case of traceless quadrupole tensor, the length scale parameters $l_i$ will be non-zero both. The physical meaning of these parameters follows from \eqref{Lo}, where one can see that $l_1$ defines the contribution of the divergence of electric field (potential part, i.e. longitudinal components of waves) to the total electric displacement, while $l_2$ defines the contribution of its rotational part (divergence-free, i.e. the transverse components).
It was shown that these "quadrupolar length" parameters are of the order of several angstroms for gases and liquids \cite{slavchov2014quadrupole, slavchov2014quadrupole2,slavchov2015polarized,lamichhane2016real,dimitrova2020quadrupolarizability}. Although for metamaterials, the magnitude of these parameters can be much larger, being determined by the characteristic size of meta-atoms and unit cells \cite{simovski2018composite}.

Let us note the following important properties of operator $L$ that directly follows from its definition \eqref{Lo}: 
\begin{equation}
\label{Lprop}
\begin{aligned}
	L\nabla f 	&= 
	(1 - l_1^2 \nabla^2) \nabla f\\
	\nabla\cdot L\textbf F 	&=   
	(1 - l_1^2 \nabla^2) \nabla\cdot\textbf F\\
	\nabla\times L\textbf F  &=  L(\nabla\times\textbf F) =
	(1 - l_2^2 \nabla^2) \nabla\times\textbf F\\
	\nabla\nabla\cdot L\textbf F 	&=   L(\nabla\nabla\cdot\textbf F) =
	(1 - l_1^2 \nabla^2) \nabla\nabla\cdot\textbf F\\
\end{aligned}
\end{equation}
where $f$ and $\textbf F$ are some arbitrary scalar and vector fields.

From \eqref{Lprop} it is seen that the application of operator $L$ to the potential and rotational fields is equivalent to application of modified Helmholtz operators with parameters $l_1$ and $l_2$ to these fields, respectively. 

By using introduced operator form of constitutive equations for $\textbf D$ field \eqref{D} we can rewrite the Maxwell equations \eqref{fe} as follows:
\begin{subequations}
\label{feL}
\begin{align}
	&\nabla\times\textbf E= -\dot{\textbf B}, \label{feLa}\\
	&\nabla\cdot\textbf B= 0, \label{feLb}\\
	&\nabla\cdot L\textbf E = \rho/\varepsilon, \label{feLc}\\
	&\nabla\times\textbf H = \textbf J+\varepsilon L\dot{\textbf E} \label{feLd}
\end{align}
\end{subequations}

The alternative form of generalized Gauss law \eqref{feLc} can be defined taking into account \eqref{Lprop}:
\begin{equation}
\label{gauss}
\begin{aligned} 
	(1 - l_1^2 \nabla^2)\nabla\cdot\textbf E = \rho/\varepsilon
\end{aligned}
\end{equation}

The form of this law \eqref{gauss} is widely used within the electrostatics with quadrupoles \cite{slavchov2014quadrupole, dimitrova2020quadrupolarizability, yue2014microscale,solyaev2021electric}.
Continuity equation for electric charge remains classical and it follows from \eqref{feLc}, \eqref{feLd}:
\begin{equation}
\label{cone}
\begin{aligned}
	\dot \rho+\nabla\cdot\textbf J =0
\end{aligned}
\end{equation}

Generalized wave equations for the vectors of electric and magnetic fields can be obtained by using \eqref{feLa}, \eqref{feLd}, constitutive relations \eqref{ce} and properties of operator $L$ \eqref{Lprop}:
\begin{equation}
\label{EH0}
\begin{aligned}
	&\nabla\times\nabla\times \textbf E= -\varepsilon\mu L\ddot{\textbf E}
	- \mu \dot{\textbf J}\\
	&\nabla\times\nabla\times\textbf B = - \varepsilon \mu L\ddot{\textbf B}
	+ \mu\nabla\times\textbf J
\end{aligned}
\end{equation}

Using standard relation for vector Laplacian together with \eqref{feLb}, \eqref{feLc} and \eqref{Lprop}, the wave equations \eqref{EH0} can be reduced to the following form:
\begin{equation}
\label{EH}
\begin{aligned}
	L(&\nabla^2\textbf E - v^{-2} L\ddot{\textbf E}) = 
			\tfrac{1}{\varepsilon}\nabla\rho+\mu L\dot{\textbf J}\\
	&\nabla^2\textbf B - v^{-2} L\ddot{\textbf B} = 
			-\mu\nabla\times\textbf J
\end{aligned}
\end{equation}
where $v = 1/\sqrt{\varepsilon\mu}$ is the speed of light in a medium. 

The obtained generalized wave equations \eqref{EH0} or \eqref{EH} can be solved directly for the vectors $\textbf E$ and $\textbf B$ by introducing appropriate hypotheses about the structure of the waves arising in a medium with quadrupole polarization (see, e.g. \cite{raab2005multipole, silveirinha2009poynting, yaghjian2016additional}). However, the general approach will involve the introduction of electromagnetic potentials \eqref{pot} by analogy with classical theory. This approach enables the consideration of the all kinds of boundary conditions of the theory \eqref{bcs}-\eqref{bcc} as well as the inherent determination of the actual wave structure (accounting, e.g. for the appearance of longitudinal components) without introducing apriori assumptions. 

%%%%%%%%%%%%%%%%%%%%%%%%%%%%%%%%%%%%%%%%%%%%%%
\section{Generalized wave equations for electromagnetic potentials and their general solution}

In this section, we present two approaches for the introduction of electromagnetic potentials and derivation of the corresponding wave equations within the multipole theory. First, we consider the isotropic material, for which the operator form of the constitutive relations is given by expression \eqref{D}. At the end of this section, we will extend our discussion to the case of general operator-based constitutive relations for $\textbf D$ field.

Substituting the definitions for electromagnetic potentials \eqref{pot} into Amp\'ere's law of multipole theory \eqref{feLd}, we obtain:
\begin{equation}
\label{gauge1}
\begin{aligned}
	&\nabla\times\textbf B = \mu\textbf J+ v^{-2} L\dot{\textbf E}
	\implies
	\nabla\times \nabla\times \textbf A = \mu\textbf J-v^{-2} L(\ddot{\textbf A}+\nabla \dot\phi)\\
\end{aligned}
\end{equation}

In this equation we can use the property of operator $L$ \eqref{Lprop} when it acts on the gradient of scalar field and also we use standard identity for the vector Laplacian, to obtain:
\begin{equation}
\label{gauge2}
\begin{aligned}
	\nabla(\nabla\cdot \textbf A 
		+ v^{-2} (1 - l_1^2 \nabla^2)\dot\phi) 
		- \nabla^2 \textbf A = \mu\textbf J - v^{-2} L\ddot{\textbf A}
\end{aligned}
\end{equation}

From \eqref{gauge2} it follows that the uncoupled equation for the vector potential can be obtained if we assume that scalar relation in brackets is zero. This provides us the modified Lorenz gauge condition for the isotropic multipole theory:
\begin{equation}
\label{ga}
\begin{aligned}
	\nabla\cdot \textbf A 
		+ v^{-2} (1 - l_1^2 \nabla^2)\dot\phi = 0 
\end{aligned}
\end{equation}
that can be reduced to classical Lorenz gauge in continuum media, when the length scale parameter has zero value, i.e. $l_1=0$.

Using \eqref{ga} in \eqref{gauge2}, we obtain the higher-order generalized wave equation for vector potential:
\begin{equation}
\label{we1}
\begin{aligned}
	v^{-2} L\ddot{\textbf A} - \nabla^2 \textbf A = \mu\textbf J 
\end{aligned}
\end{equation}

This equation can be further simplified by using Helmholtz decomposition theorem for the vector potential and for the current density:
\begin{equation}
\label{helm}
\begin{aligned}
	\textbf A &= \pmb\Psi+\nabla\psi, \quad \nabla\cdot\pmb\Psi=0\\
	\textbf J &= \textbf J_s + \nabla J_p, \quad \nabla \cdot\textbf J_s=0 
\end{aligned}
\end{equation}
where we introduce auxiliary potentials $\pmb\Psi$ and $\psi$ that define the solenoidal and potential (longitudinal) parts of vector potential $\textbf A$, respectively. Note that in the general case of electrodynamic processes in continuum media these parts can be non-zero. The current density is also represented as the sum of its solenoidal part $\textbf J_s$ and potential part defined via scalar field $J_p$, for which the continuity equation \eqref{cone} yields: $\nabla^2 J_p=-\dot\rho$.

Using \eqref{helm}, we can split the general equation \eqref{we1} into two separate wave equations for the solenoidal and potential parts of $\textbf A$ field. For these parts we can use the properties of operator $L$ \eqref{Lprop} and obtain the following simplified equations:
\begin{equation}
\label{we12}
\begin{aligned}
	&v^{-2} (1-l_2^2\nabla^2)\ddot{\pmb \Psi} 
		- \nabla^2 \pmb \Psi = \mu\textbf J_s\\
	&v^{-2} (1-l_1^2\nabla^2)\ddot{\psi} 
		- \nabla^2 \psi = \mu J_p
\end{aligned}
\end{equation}

Taking into account \eqref{helm}, the gauge condition \eqref{ga} is reduced to the following form: 
\begin{equation}
\label{ga2}
\begin{aligned}
	\nabla^2\psi
		+ v^{-2} (1 - l_1^2 \nabla^2)\dot\phi = 0 
\end{aligned}
\end{equation}

The generalized wave equation for the scalar potential can be established by substituting \eqref{pot} into Gauss's law \eqref{feLc}, yielding:
\begin{equation}
\label{we20}
\begin{aligned}
	% &(1-l_1^2\nabla^2)\nabla\cdot\textbf E = \rho/\varepsilon
	% \implies 
	(1-l_1^2\nabla^2)\nabla\cdot(-\dot{\textbf A}-\nabla\phi) = \rho/\varepsilon
\end{aligned}
\end{equation}
in which we can use the gauge condition \eqref{ga} and finally obtain:
\begin{equation}
\label{we2}
\begin{aligned}
	(1-l_1^2\nabla^2)(v^{-2}(1-l_1^2\nabla^2) \ddot\phi-\nabla^2\phi) = \rho/\varepsilon
\end{aligned}
\end{equation}

Thus, within the considered variant of multipole theory with electric quadrupole-magnetic dipole approximation, the spatial order of wave equation for vector potential and for $\textbf B$ field remains classical, while for scalar potential and for $\textbf E$ field it has the fourth order (see \eqref{EH}, \eqref{we12}, \eqref{we2}). In the higher-order multipole theories the order of corresponding wave equations will increase.

Considering time-harmonic processes and assuming that all field variables varies in time as $\sim \text e^{-\text i \omega t}$, we obtain the following form of wave equations \eqref{we12}, \eqref{we2}:
\begin{equation}
\label{th}
\begin{aligned}
	&(k_2^2 + \nabla^2) \pmb \Psi = 
			-\mu s^2_2\textbf J_s\\
	&(k_1^2 + \nabla^2) \psi = 
			-\mu s^2_1 J_p\\
	&(1-l_1^2\nabla^2)(k_1^2 + \nabla^2)\phi = 
			-s^2_1\rho/\varepsilon
\end{aligned}
\end{equation}
where $k_i^2 = v^{-2} \omega^{2} s_i^2$ ($i=1,2$) are the wavenumbers, in which we introduce the definition for frequency-dependent coefficients $s_i = 1/\sqrt{1 - l_i^2v^{-2} \omega^{2}} = \sqrt{1+l_i^2 k_i^2}$ (no summation over $i$) for convenience of the following derivations.

It can be seen, that system of wave equations \eqref{th} will be reduced to classical one in the case $l_i=0$ (absence of quadrupole effects), for which we have $s_1=s_2=1$ and the definitions for wavenumbers reduces to single relation $k_1^2 = k_2^2 = v^{-2} \omega^{2}$. In general case of  multipole theory ($l_1\neq l_2$ that is also the case for traceless $\textbf Q$), the values of $k_1$ and $k_2$ are different. 

The representation of general solution for wave equations \eqref{th} can be given in the following form:
\begin{equation}
\label{gs}
\begin{aligned}
	\pmb\Psi &= \tilde{\pmb\Psi} + \bar{\pmb\Psi}, \quad
	\mathcal H_2\tilde{\pmb\Psi} =0,\quad
	\mathcal H_2\bar{\pmb\Psi} =-\mu s^2_2\textbf J_s,\quad\\
	\psi &= \tilde{\psi} + \bar{\psi}, \quad
	\mathcal H_1\tilde{\psi} =0,\quad 
	\mathcal H_1\bar{\psi} = -\mu s^2_1 J_p,\quad\\
	\phi &= \tilde{\phi} + \hat{\phi} + \bar\phi, \quad
	\mathcal H_1\tilde{\phi} =0,\quad 
	\mathcal M_1\hat{\phi} =0,\quad
	\mathcal M_1\mathcal H_1\bar\phi = -\tfrac{s^2_1}{\varepsilon}\rho\\
\end{aligned}
\end{equation}
where we use the notation for Helmholtz operators $\mathcal H_i = k_i^2 + \nabla^2$ ($i=1,2$) and modified Helmholtz operator $\mathcal M_1 = 1- l_1^2\nabla^2$. The tilde symbol " $\tilde {}$ " defines the propagated part of the wave that obey the homogeneous Helmholtz equation. The hat symbol " $\hat{}$ "defines the evanescent wave that obey the homogeneous modified Helmholtz equation. The bar symbol "$\bar{ }$ " defines the particular solutions related to the volume density of currents $\textbf J$ and charges $\rho$.

Therefore, the general solution for the potentials $\pmb\Psi$ and $\psi$ in \eqref{gs} retains its classical form, where only the definitions for wavenumber $k_i$ and source terms are modified, while the solution structure itself remains unchanged and corresponds to Helmholtz equation. In contrast, the governing equation for the scalar potential $\phi$ undergoes a structural change, leading to the appearance of an additional evanescent component. The ability to express the general solution for $\phi$ as a sum of a propagating wave $\tilde\phi$ and an evanescent  wave $\hat\phi$ stems from the fact that its governing equation is formed by a composition of two commuting operators $\mathcal H_1$ and $\mathcal M_1$ (see \eqref{th}) \cite{lazar2014gradient, lazar2020second}.

Using representation \eqref{gs} in \eqref{ga2} and taking into account definitions for $k_i$, one can find that the modified Lorenz gauge is reduced to a simple proportionality between the propagating parts of scalar potentials:
\begin{equation}
\label{gauge}
\begin{aligned}
	\tilde\phi= -\text i\omega\tilde\psi
\end{aligned}
\end{equation}
 while the consistency of the particular solutions is ensured by satisfying the continuity equation \eqref{cone}, and the evanescent wave $\hat\phi$ is eliminated from \eqref{gauge} due to the presence of the operator $\mathcal M_1$ in relation \eqref{ga2}.

Using representation \eqref{gs} and taking into account \eqref{helm}, \eqref{gauge} in \eqref{pot}, we obtain the following definitions for electric and magnetic fields via introduced potentials for the time-harmonic processes:
\begin{equation}
\label{defE}
\begin{aligned}
	\textbf E	&= \text i \omega\tilde{\pmb\Psi}
				-\nabla(2 \tilde\phi + \hat\phi) + \bar{\textbf E}
\end{aligned}
\end{equation}
\begin{equation}
\label{defB}
\begin{aligned}
	\textbf B	&=  \nabla\times\tilde{\pmb\Psi} + \bar{\textbf B}
\end{aligned}
\end{equation}
in which the particular solutions $\bar{\textbf E}$ and $\bar{\textbf B}$ are related to the corresponding particular solutions for  potentials $\bar{\pmb\Psi}$, $\bar \psi$, $\bar\phi$ in \eqref{gs}. Considering relations \eqref{pot} and the governing equations for the particular solutions \eqref{gs}, we arrive at:
\begin{equation}
\label{defEx}
\begin{aligned}
	\bar{\textbf E}&= \text i \omega\bar{\pmb\Psi}
				+\text i \omega\nabla\bar{\psi}
				-\nabla\bar\phi
			   = \text i\mu s^2_2 (G_2\ast\textbf J_s)
				+\text i\mu s^2_1 \nabla(G_1\ast J_p)
				+\tfrac{s^2_1}{\varepsilon} \nabla(G_{11}\ast\rho)\\[5pt]
	\bar{\textbf B}&= \nabla\times \bar{\pmb \Psi}
			   = - \mu s_2^2\, 
			   \nabla\times(G_2\ast \textbf J_s)
\end{aligned}
\end{equation}
where $G_i = G_i(k_i)$ are the Green functions of Helmholtz operators $\mathcal H_i$ that depend on the wavenumber $k_i = \omega s_i/v$ (introduced above); symbol "$\ast$" defines convolution; and $G_{11}$ is the Green function of the operators composition $\mathcal M_1 \mathcal H_1$, which can be derived by using Fourier transform in the following form:
\begin{equation}
\label{G11}
\begin{aligned}
	G_{11}(k_1, l_1) = \frac{1}{1+l^2_1k_1^2}
	\left(
		G_1(k_1) - G_1(\text i/l_1)
	\right)
\end{aligned}
\end{equation}

The obtained representations for particular solutions \eqref{defEx} are essentially a generalization of the classical dyadic Green's functions for the electric and magnetic fields within the considered multipole theory. In the absence of quadrupole effects ($l_i=0$, $k_1=k_2=k = \omega/v$), these expressions reduce to the classical formulations, which are expressed solely in terms of the Green's functions for the Helmholtz operator $G_1(k)$. 
Expressions for the dyadic Green's functions \eqref{defEx} in the particular problems of multipole theory in 1D, 2D, 3D can be derived by using  corresponding well-known representations for the Green's function of  Helmholtz equation $G_i(k_i)$ \cite{morse1946methods}.

The general solution representation derived for multipole theory \eqref{gs}-\eqref{G11} is new and has not been examined before to the author's knowledge. This formulation enables convenient analysis of the solutions  for boundary value problems in multipole theory, which will be illustrated further in the context of wave propagation and reflection problems. 

Concluding this section, we show that the proposed approach for constructing the general solution can be extended to more complex forms of the operator appearing in the constitutive relations for electric displacement $\textbf D$ field \eqref{D}, \eqref{Lo}. Namely, let us consider the constitutive equations $\textbf D = \varepsilon L \textbf E$, in which $L$ is arbitrary  linear differential operator. Using this relation together with definition \eqref{pot} in the Amp\'ere's law \eqref{feLd}, we obtain:
\begin{equation}
\label{gaugeL}
\begin{aligned}
	\left(\nabla\nabla\cdot \textbf A 
		+ v^{-2} L\nabla\dot\phi\right) 
		- \nabla^2 \textbf A = \mu\textbf J - v^{-2} L\ddot{\textbf A}
\end{aligned}
\end{equation}

Now, the relation in the brackets cannot be considered as the gauge condition for the potentials, since this relation is vector. In general, we do not know how the operator $L$ acts on the gradient of a scalar function ($\nabla\dot\phi$), and it is impossible to factor this gradient out of the expression.
Nevertheless, we can introduce the auxiliary vector potential $\textbf A_{aux}$, to define the electric vector potential $\textbf A$ in the following form:
\begin{equation}
\label{gaugeL2}
\begin{aligned}
	\textbf A = L \textbf A_{aux}
\end{aligned}
\end{equation}

Note that definition \eqref{gaugeL2} is sufficiently general provided that the action of operator $L$ on a vector field preserves its potential and solenoidal components. Otherwise, we could immediately utilize this property of operator $L$ to obtain the required gauge form in expression \eqref{gaugeL}. By substituting expression \eqref{gaugeL2} into \eqref{gaugeL}  under the assumption that operator $L$ commutes with the $"\nabla\nabla\cdot$" operator (an assumption that restricts the method's generality but is justifiable for numerous types of constitutive relations) we derive another version of the modified Lorenz gauge condition and corresponding uncoupled generalized wave equation expressed in terms of the auxiliary potential:
\begin{equation}
\label{gaugeL3}
\begin{aligned}
	\nabla\cdot \textbf A_{aux} 
		+ v^{-2} \nabla\dot\phi = 0, \qquad 
		v^{-2} L\ddot{\textbf A}_{aux} - \nabla^2 \textbf A_{aux} = \mu\textbf J 
\end{aligned}
\end{equation}

The general solution of wave equation in \eqref{gaugeL3} together with that for the scalar potential \eqref{we2}, can be constructed in a manner analogous to the approach used in \eqref{gs}-\eqref{G11}. The electric vector potential $\textbf{A}$ and the electromagnetic fields $\textbf{E}$ and $\textbf{B}$ can then be expressed in terms of $\textbf{A}_{\text{aux}}$ and $\phi$ through definitions \eqref{pot} and \eqref{gaugeL2}. This framework enables the solution of boundary value problems within the multipole theory \eqref{fe}-\eqref{bce} with more general constitutive equations, including the models featuring magnetic and anisotropic effects as well as the models with higher-order multipole effects. For the last one, the operator $L$ will contain additional higher-order derivatives \cite{raab2005multipole, kafadar1971theory}.

%%%%%%%%%%%%%%%%%%%%%%%%%%%%%%%%%%%%%%%%%%%%%%
\section{Propagation phenomena}

Let us consider a propagation of plane electromagnetic wave in an infinite  dielectric media with quadrupole effects in absence of free charges and currents.
Solution for the field variables is found in the form $\sim \text e^{\text i (\textbf k\cdot\textbf x-\omega t)}$ ($\textbf k$ is the wave vector, $\omega$ is angular frequency). From Maxwell equations \eqref{fe} we have as usual
\begin{equation}
\label{propME}
\begin{aligned}
	\textbf k\times \textbf E = \omega \textbf B, \qquad
	\textbf k\cdot \textbf B = 0, \qquad
	\textbf k\cdot \textbf D = 0, \qquad
	\textbf k\times \textbf H = -\omega \textbf D
\end{aligned}
\end{equation}

Thus, the magnetic and electric displacement fields are always transverse to the wave vector. However, the directions of $\textbf D$ and $\textbf E$ may not be the same due to the presence of weak spatial dispersions \cite{landau2013electrodynamics,agranovich1966spatial} and the electric field $\textbf E$ may have the transverse and longitudinal components both. These components of $\textbf E$ are defined via the solenoidal field $\tilde{\pmb \Psi}$ and scalar potential $\tilde\phi$ \eqref{defE} and they are governed by Helmholtz equations $\mathcal H_1$ and $\mathcal H_2$ \eqref{th}, respectively. These equations have different definitions for wavenumbers $k_i$ (see \eqref{th}) and they provide us the following two kinds of dispersion relations:
\begin{equation}
\label{prop2}
\begin{aligned}
	k_i = \frac{\omega s_i}{v}
	 = \frac{\omega}{v\sqrt{1-l_i^2\omega^2v^{-2}}} 
	 = \frac{\omega \hat n_i}{c} \qquad (i=1,2)
\end{aligned}
\end{equation}
where coefficients $s_i$ were defined above in \eqref{th} and we introduce the notation for the effective refractive indices $\hat n_i$ that takes into account the spatial dispersion effects:
\begin{equation}
\label{hn}
\begin{aligned}
\hat n_i= n s_i=\frac{n}{\sqrt{1-l_i^2\omega^2v^{-2}}} = n\sqrt{1+l_i^2 k_i^2}
\end{aligned}
\end{equation}
and we use standard definition $n=c/v=c\sqrt{\varepsilon\mu}$ for the limiting value of $\hat n$ in the low-frequency/long-wavelength approximation ($\omega\rightarrow0,\,k_i\rightarrow0$), when the quadrupolar polarization effects become negligible.

Therefore, in a quadrupolar medium, the longitudinal component of the electromagnetic wave will exhibit distinct phase velocity and effective refractive index compared to transverse components. This distinction arises because the length scale parameters $l_1$ and $l_2$ are generally unequal. While dispersion relations for longitudinal and transverse waves coincide in the classical long-wavelength limit (i.e., when the wavelengths significantly exceed the length scale parameters), their behavior diverges increasingly at higher frequencies. Thus, the theoretical prediction indicates that longitudinal waves should become particularly prominent and observable in high-frequency regimes in quadrupolar media (in metamaterials, in quadrupolar liquid mixtures, etc.). The evanescent longitudinal component $\hat\phi$ does not arise in the infinite media since it is localised around the boundaries and ihomogeneities that is the property of its governing operator $\mathcal M_1$ \eqref{gs}. 

\begin{figure}[t!]
 \centering
  (a)\includegraphics[width=0.4\linewidth]{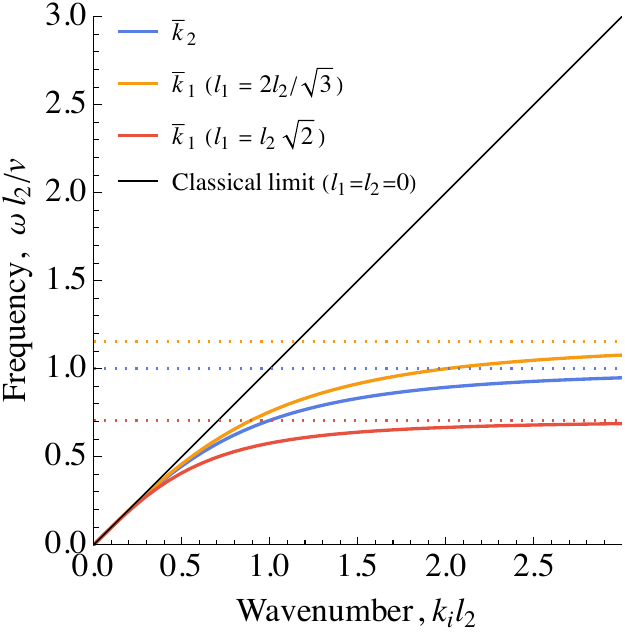}\quad
  (b)\includegraphics[width=0.4\linewidth]{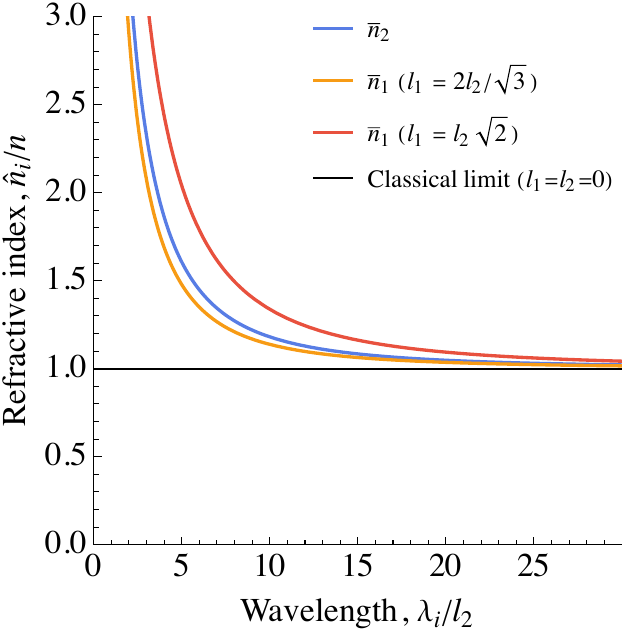}
  \caption{(a): Dispersion relations in the media with weak spatial dispersion for transverse ($k_2$) and longitudinal ($k_1$) components of electromagnetic wave for different values of the length scale parameters. Horizontal dotted lines correspond to the asymptotic cutoff frequencies $\omega_c = v/l_i$. (b): Dependence of effective refractive indices $\hat n_i$ on wavelength $\lambda_i=2\pi/k_i$.}
  \label{fig1}
\end{figure}

Note that the form of dispersion relations \eqref{prop2} for the transverse waves have been previously discussed in works \cite{kafadar1971theory, simovski2018composite, serdukov2001electromagnetics} without consideration of longitudinal components. The illustrations for these relations for the case of traceless tensor of quadrupoles ($\alpha_1=-2\alpha_2/3$, $l_1=2l_2/\sqrt{3}$) \cite{silveirinha2014boundary,yaghjian2020comment} and for the simplified model of quadrupolar media ($\alpha_1=0$, $l_1=\sqrt{2} l_2$) \cite{simovski2018composite} are presented in Fig. \ref{fig1}. It can be observed that the medium exhibits normal spatial dispersion, and for short wavelengths, an increase in the effective refractive index occurs. At large wavenumber, the wave frequency approaches asymptotic value (cutoff frequency) determined by the ratio of the wave velocity to the corresponding length scale parameter $\omega_c = v/l_i$ (see dotted lines in Fig. \ref{fig1}a). The effective refractive index of the longitudinal wave component $\hat n_1$ can be either higher or lower than that of the transverse components $\hat n_2$, depending on the ratio $l_1/l_2$ (Fig. \ref{fig1}b). For the most common case of a medium with a traceless quadrupole polarization tensor, the longitudinal component propagates with a higher phase velocity than the transverse waves, as $\hat n_1<\hat n_2$ (see orange line in Fig. \ref{fig1}b).

\newpage

By using \eqref{propME}, \eqref{prop2} we can find the wave impedance and use it to give explicit definition for Poynting vector within the considered multipolar theory. Without loss of generality, we can assume that wave propagates along $z$ axis direction and that the electromagnetic fields can be defined as follows:
\begin{equation}
\label{propf}
\begin{aligned}
	\textbf E &= E_{\perp} \text e^{\text i ( k_2 z-\omega t)}\textbf e_x + E_{\|} \text e^{\text i ( k_1 z-\omega t)}\textbf e_z,\\
	\textbf H &= H \text e^{\text i ( k_2 z-\omega t)}\textbf e_y
\end{aligned}
\end{equation}
where $E_{\perp}$ and $E_{\|}$ are the transverse and longitudinal components of electric field that can be related to the potentials $\pmb \Psi$ and $\phi$ used above in \eqref{defE}.

The tensor of electric quadrupoles \eqref{ceq} can be presented then in the following form: 
\begin{equation}
\label{propq}
\begin{aligned}
	\textbf Q& = \text i k_2 \l_2^2\varepsilon E_{\perp}e^{\text i ( k_2 z-\omega t)}(\textbf e_x\otimes\textbf e_z+\textbf e_z\otimes\textbf e_x)
	+ \text i k_1 \l_1^2\varepsilon E_\|e^{\text i ( k_1 z-\omega t)}\textbf e_z\otimes\textbf e_z
\end{aligned}
\end{equation}

The relation between the amplitudes of magnetic and electric fields can be found from the first equation in \eqref{propME} and it is reduced to:
\begin{equation}
\label{imp0}
\begin{aligned}
	&|\textbf k\times \textbf E| = \omega \mu |\textbf H|\implies
	k E_\perp = \omega \mu H
\end{aligned}
\end{equation}

Using dispersion relations \eqref{prop2} in \eqref{imp0}, we find:
\begin{equation}
\label{imp}
\begin{aligned}
	&H = \frac{\hat n_2}{c \mu}E_\perp 
		= \sqrt{\frac{\varepsilon}{\mu}} s_2 E_\perp = \frac{E_\perp s_2}{Z}
\end{aligned}
\end{equation}
where $Z = \sqrt{\mu/\varepsilon}$ is the wave impedance.

Within the presented theory, the Poynting vector for a propagating plane wave must be calculated using a definition that accounts for spatial dispersion \cite{yaghjian2016additional, yaghjian2018power}. This expression can also be obtained directly from Maxwell's equations within the derivation of Poynting theorem (see Appendix B). The resulting definitions for the Poynting vector and its time-averaged value for time-harmonic fields are:
\begin{equation}
\label{Sg}
\begin{aligned}
	\textbf S &= \textbf E\times \textbf H - \dot{\textbf Q}\cdot\textbf E\\
	\langle\textbf S\rangle &=  
	\frac{1}{2}\,\text{Re}(\textbf E\times\textbf H^* 
		+ \text i \omega \,\textbf E^*\cdot\textbf Q)
\end{aligned}
\end{equation}
where the asterisk superscript denotes the complex conjugate.

Using \eqref{propf}, \eqref{propq}, \eqref{imp} and \eqref{prop2} in \eqref{Sg}  one can obtain the following general definition for the magnitude of energy flux (for derivations, see Appendix B):
\begin{equation}
\label{Sf}
\begin{aligned}
	S = |\langle\textbf S\rangle| = \frac{1}{2}\left(
	\frac{E_\perp^2}{Z s_2} - \frac{E_\|^2}{Z s_1}k_1^2 l_1^2
	\right)
\end{aligned}
\end{equation} 
% where we take into account that the electromagnetic wave may have transverse ($E_\perp$) and longitudinal ($E_\|$) components of electric field in the media with quadrupole effects.

Note that if we use the classical definition of Poynting vector instead of \eqref{Sg}, \eqref{Sf} (as employed, for instance, in \cite{de2006surprises, achouri2021extension}), the resulting expression for $S$ not only fails to account for the longitudinal field component $E_\|$ but also inaccurately represents the contribution of the transverse component:
\begin{equation}
\label{Sgi}
\begin{aligned}
	\textbf S &= \textbf E\times \textbf H \implies
	\langle\textbf S\rangle =  
	\frac{1}{2}(\textbf E\times\textbf H^*) \implies
	S = \frac{E_\perp^2 s_2}{2Z}
\end{aligned}
\end{equation}

Using this expression \eqref{Sgi} within the multipole theory makes it impossible to satisfy the energy balance even for purely transverse waves. We will illustrate this fact explicitly in the next section within a closed-form analytical solution for the normal incidence problem.

%%%%%%%%%%%%%%%%%%%%%%%%%%%%%%%%%%%%%%%%%%%%%%
\section{Reflection phenomena}

\begin{figure}[h!]
 \centering
  \includegraphics[width=0.75\linewidth]{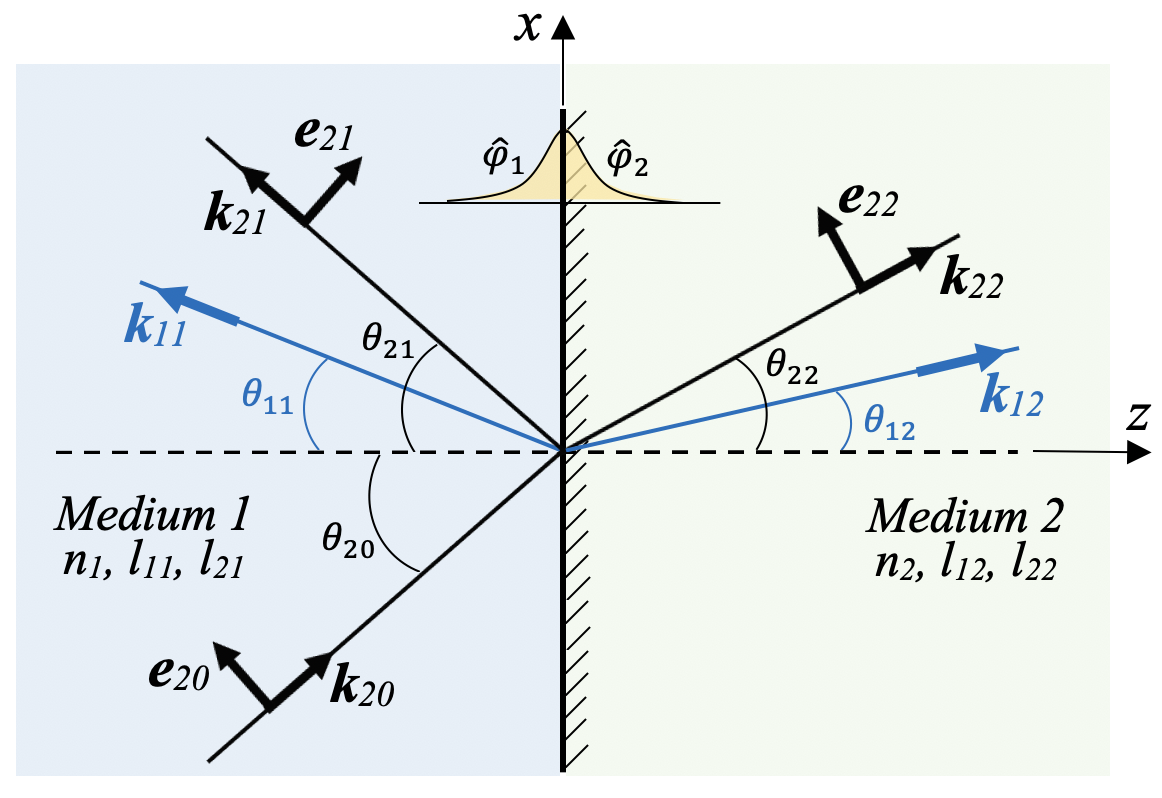}
  \caption{Plane wave incidence on a planar interface. Propagation directions defined by wave vectors $\textbf k_{ij}$ correspond to transverse waves (black lines) and to the longitudinal components of reflected and transmitted waves (blue lines); $\textbf e_{ij}$ -- polarization unit vectors, $\hat \phi_j$ -- potentials of evanescent waves. Angle of incidence is $\theta_{20}$.}
  \label{fig2}
\end{figure}

Consider a plane wave propagating in a medium with refractive index $n_1 = c/v_1=c\sqrt{\varepsilon_1\mu_1}$, which is incident on a  plane boundary of a medium with refractive index $n_2 = c/v_2=c\sqrt{\varepsilon_2\mu_2}$ (Fig. \ref{fig2}). 
Both media exhibit dipole and quadrupole electric polarization and they are non-magnetic ($\mu_1=\mu_2=\mu_0$). Both length scale parameters are non-zero in the media, and we will denote them as $l_{ij}$, where $i=1,2$ indicates the parameter index in the constitutive relations \eqref{D}, \eqref{Lo}, and $j=1,2$ denotes the media number. 
The incident wave is assumed to be transverse, while the structure of the reflected and transmitted waves is found from the solution. All processes are time-harmonic and we will omit factor $\text e^{-\text i\omega t}$ in all subsequent expressions. 

%%%%%%%%%%%%%%%%%%%%%%%%%%%%%%%%%%%%%%%%%%%%%%
\subsection{Normal incidence}

Initially, we examine the case of normal incidence ($\theta_{20}=0$). It can be easily shown that in this case, the set of non-trivial continuity conditions in quadrupolar media \eqref{bcc} reduces to the continuity of the tangential components of the magnetic and electric fields only (similar to classical solution). The remaining boundary conditions are either trivial, non-independent, or reduce to the requirement of the absence of the longitudinal electric field component in the reflected and transmitted waves. Thus, the solution for scalar electric potentials in both materials is $\phi_j  = 0$ ($j=1,2$), while for the transverse field we can define the vector potentials according to \eqref{helm}, \eqref{gs} in the following form:
\begin{equation}
\label{gsA}
\begin{aligned}
	\textbf A_j(z,t) &
	= \Psi_j(z,t) \textbf e _x,
	\qquad (k_{2j} + \nabla^2) \Psi_j =0
\end{aligned}
\end{equation}
where index $j=1,2$ defines the media number; without loss of generality we  assume that wave is polarized along $x$-axis; $\Psi_j$ are the projections of vector potentials on $x$-axis (we omit tilde symbol for these potentials); $k_{2j} = \omega \hat n_{2j}/c$ are the wavenumbers defined according to the solution of propagation problem for the transverse waves \eqref{prop2}, $\hat n_{2j} = n_j s_{2j}$ are the effective refractive indices for transverse waves \eqref{hn}, and 
$s_{2j}= 1/\sqrt{1-l_{2j}^2\omega^2v^{-2}_j}$ are the coefficients that define the non-classical frequency-dependent effects. 

The functions $\Psi_j$ can be defined then as follows:
\begin{equation}
\label{potn}
\begin{aligned}
	z<0:
	\Psi_1 &= A_0 \text e^{\text i k_{21} z} 
				+ A_1 \text e^{-\text i k_{21} z} \\
	z>0:
	\Psi_2 &= A_2 \text e^{\text i k_{22} z} 
\end{aligned}
\end{equation}
where $A_0$, $A_1$ and $A_2$ are the amplitudes of vector potential of incident, reflected and transmitted waves, respectively; and we take into account that $\textbf k_{21}= - \textbf k_{20}$, $k_{21}=|\textbf k_{21}|=k_{20}$ that follows from the homogeneity condition of the solution along $x$-axis.

By using \eqref{gsA}, \eqref{potn} in \eqref{pot}, \eqref{ceq}, \eqref{D} and taking into account \eqref{Lprop}, we find the field variables:
\begin{equation}
\label{gsn}
\begin{aligned}
	\textbf E_j&= \text i\omega\textbf A_j=  \text i\omega\Psi_j \textbf e _x\\
	\textbf B_j&=\mu\textbf H_j=\nabla\times\textbf A_j = \Psi'_j \textbf e _y\\
	\textbf Q_j&= \alpha_{2j} (\nabla\textbf E_j + \textbf E_j\nabla) = 
	\text i \omega l_{2j}^2\varepsilon_j\Psi'_j(\textbf e _x\otimes\textbf e _z+\textbf e _z\otimes\textbf e _x)\\
	\textbf D_j&= \varepsilon L \textbf E_j = 
		\text i \omega\varepsilon_j 
		 s_{2j}^2\Psi_j\textbf e _x
\end{aligned}
\end{equation}
where $j=1,2$ is the media number and prime symbol defines derivative $d/d z$.

Using \eqref{gsn} in \eqref{bcc}, one can obtain the following representation for the non-trivial independent continuity conditions at the boundary $z=0$:
\begin{equation}
\label{BCnc}
\begin{aligned}
	&[\textbf n\times\textbf H
	-\text i\omega\,\textbf n\cdot\textbf Q\cdot(\textbf I - \textbf n\otimes\textbf n)]=0
	\implies
	[H_y + \text i\omega Q_{zx}]=0
	\implies 
	[s_{2j}^{-2} \Psi'_j]=0 
	\\
	&[\textbf n\times\textbf E] = 0
	\implies 
	[E_x] = 0
	\implies
	\,[\Psi_j]=0
\end{aligned}
\end{equation}

Substituting \eqref{potn} into \eqref{BCnc}, we obtain the following system and solution for amplitudes $A_i$:
\begin{equation}
\label{sysn}
\begin{aligned}
	&\frac{\hat n_2}{s^2_{22}}A_2 = \frac{\hat n_1}{s^2_{21}}(A_0 - A_1),\quad
	A_2 = A_0 + A_1
	 \,\,\implies \,\,
	 A_1 = \frac{1 -  \hat m}{1+ \hat m}A_0,\quad 
	 A_2 = \frac{2}{1+ \hat m}  A_0
\end{aligned}
\end{equation}
where we introduce coefficient $\hat m = \frac{\hat n_{22} s^2_{21}}{\hat n_{21} s^2_{22}} = \frac{n_2 s_{21}}{n_1 s_{22}} = m \frac{s_{21}}{s_{22}}$ that has the meaning of effective relative refractive index of media 2 with respect to the media 1 within the multipole theory. This coefficient reduces to classical value $\hat m = m=n_2/n_1$ when $l_{2j}\rightarrow 0$, i.e. when the quadrupole effects are negligible.

Based on \eqref{gsn}, \eqref{sysn} we can introduce the explicit representations for the reflection and transmission coefficients:
\begin{equation}
\label{soln}
\begin{aligned}
	\tilde r = \frac{|\textbf E_1|}{|\textbf E_0|} 
			= \frac{1 - \hat m}{1+ \hat m},\qquad
	\tilde t = \frac{|\textbf E_2|}{|\textbf E_0|} =\frac{2}{1+\hat m}
\end{aligned}
\end{equation}
that take classical form in terms of effective coefficient $\hat m$ (see, e.g. \cite{bohren2008absorption}) .

The reflectance $R$ and transmittance $T$ coefficients can be defined then as the ratios between the amplitudes of corresponding energy fluxes taking into account \eqref{soln} and definition for Poynting vector \eqref{Sf}:
 
\begin{equation}
\label{enn}
\begin{aligned}
	R &= \frac{S_1}{S_0} 
	=  \frac{|\textbf E_1|^2}{|\textbf E_0|^2} = \tilde r^2\\
	T &= \frac{S_2}{S_0} = \frac{|\textbf E_2|^2}{|\textbf E_0|^2}\frac{Z_1 s_{21}}{Z_2 s_{22}}
	= \tilde t^2 \sqrt{\frac{\varepsilon_2}{\varepsilon_1}}\frac{s_{21}}{s_{22}}
	= \tilde t^2 \frac{n_2}{n_1}\frac{s_{21}}{s_{22}} = \tilde t^2 \hat m
\end{aligned}
\end{equation}
 that also have classical form up to definition of coefficient $\hat m$ \cite{bohren2008absorption}.
 
 From \eqref{soln}, \eqref{enn} it is easy to see that $T+R=1$, i.e.  the energy balance condition is preserved. 
 However, if we try to use classical definition for Poynting vector \eqref{Sgi} in \eqref{enn}, then we obtain the incorrect condition $T+R = 1- \tfrac{4m \bar s (1 - \bar s^2)}{(m+\bar s)^2}$, where $\bar s = s_{22}/s_{21}$.
 
 Thus, the presented solution demonstrates that the correct definition for Poynting vector must take into account the contribution of quadrupole effects \eqref{Sf}. 
Previously, this closed-form solution had not been derived within the multipole theory. Consequently, some of existing studies utilized the classical Poynting vector definition, introducing artifacts in energy balance as discussed in the Introduction.

\begin{figure}[b!]
 \centering
  (a)\includegraphics[width=0.4\linewidth]{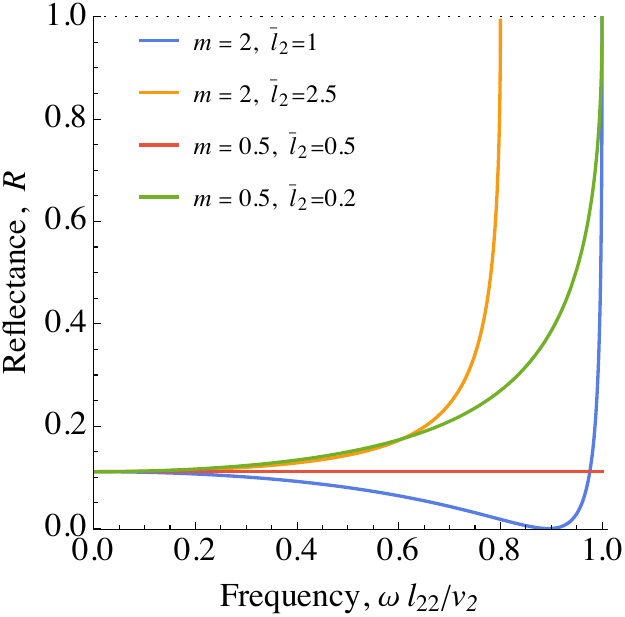}\quad
  (b)\includegraphics[width=0.4\linewidth]{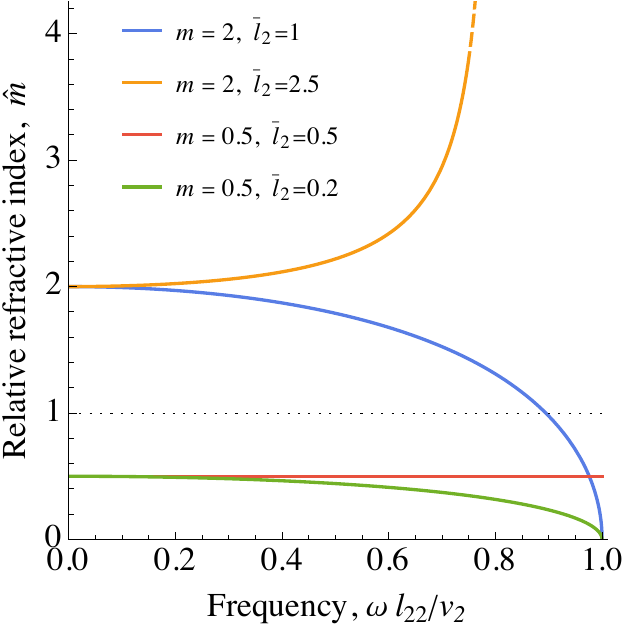}
  \caption{(a): Dependence of reflectance coefficient $R$ (a) and effective relative refractive index $\hat m$ (b) on frequency in the solution of normal incidence problem for the media with weak spatial dispersion.}
  \label{fig3}
\end{figure}

Illustration for the established dependence of reflectance coefficient $R$ \eqref{enn} and effective relative refractive index $\hat m$ on frequency is given in Fig. \ref{fig3} for several variants of the ratios between the refractive indices $m=n_2/n_1$ and the length scale parameters $\bar l_2 = l_{21}/l_{22}$.
It can be observed in Fig. \ref{fig3} that at low frequencies, $R$ and $\hat m$ take their classical values. As the frequency increases, non-classical effects emerge, whose specific character is determined by the values of material constants. When the ratios $m$ and $\bar l_2$ are equal, no non-classical effects occur (see red lines in Fig. \ref{fig3}). However, in all other cases, the model predicts a change of the reflectance coefficient that strongly increases in the high-frequency range. Once the incident wave frequency reaches the cutoff frequency of the second layer ($\omega \rightarrow \omega_c=v_2/l_{22}$, see Fig. \ref{fig1}), total internal reflection is realized. This corresponds to the wave's inability to propagate through the frequency band gap of the second material. In this case, the relative refractive index $\hat m$ becomes either zero or infinite (Fig. \ref{fig3}b), which, in either scenario, results in a reflection coefficient of $R=1$ in the obtained solution (see \eqref{soln}, \eqref{enn}).

Notably, the increase in the reflectance coefficient $R$ can be either monotonic or non-monotonic depending on the values of $m$ and $\bar l_2$. In the latter case, there may exist a specific frequency at which the total transmission effect occurs (see blue lines in Fig. \ref{fig2}). This condition corresponds to the frequency where the effective relative refractive index becomes unity ($\hat m =1$).

%%%%%%%%%%%%%%%%%%%%%%%%%%%%%%%%%%%%%%%%%%%%%%
\subsection{Oblique incidence}

We now proceed to analyze the oblique incidence of a plane wave with the electric field vector to be parallel to the plane of incidence ($xz$). This configuration serves as a clear example for demonstrating the features of the solution that accounts for all boundary conditions of the multipole theory \eqref{bcc}. 
% The problem of oblique incidence with wave polarization perpendicular to the plane of incidence is remained for future study.

From definitions \eqref{defE}, \eqref{defB} it follows, that the reflected and transmitted waves may have a propagated and an evanescent longitudinal components defined by potentials $\tilde \phi$ and $\hat \phi$, respectively. The problem thus requires characterization of three distinct wave types in both materials: propagating transverse waves, propagating longitudinal waves, and evanescent longitudinal waves. This yields six waves in total, contrasting with the classical case of isotropic media where only two purely transverse waves (reflected and transmitted) emerge.

Generally, transverse and longitudinal components are characterized by distinct wave vectors that will be denoted as $\textbf k_{ij}$, where, the first index 
$i=1,2$ indicates the component type (longitudinal or transverse), while the second index 
$j=0...3$ identifies the wave type (0 -- incident, 1 -- reflected, 2 -- transmitted). 
Therefore, for purely transverse incident wave we define (see Fig. \ref{fig2}):
\begin{equation}
\label{k20}
\begin{aligned}
	\textbf k_{20} &= k_{20}(\sin \theta_{20} \,\textbf e_x + \cos \theta_{20} \,\textbf e_z)
\end{aligned}
\end{equation}
where $\theta_{20}$ is the prescribed angle of incidence, $k_{20} = |\textbf k_{20}| = \omega \hat n_{21}/c$ is the wavenumber of incident wave, $\hat n_{21} = n_1 s_{21}$ is the effective refractive index for incident wave, which spatial dispersion is defined by the coefficient $s_{21}= 1/\sqrt{1-l_{21}^2\omega^2v^{-2}_1}$ (see \eqref{prop2}, \eqref{hn}).

For the propagating reflected and transmitted waves, we note that the polarization plane is preserved in the isotropic material and define (see Fig. \ref{fig2}):
\begin{equation}
\label{kij}
\begin{aligned}
	\textbf k_{i1} &= k_{i1}(\sin \theta_{i1} \,\textbf e_x - \cos \theta_{i1} \,\textbf e_z),\qquad
	\textbf k_{i2} = k_{i2}(\sin \theta_{i2} \,\textbf e_x + \cos \theta_{i2} \,\textbf e_z)
\end{aligned}
\end{equation}
where $i=1, 2$ corresponds to the longitudinal and transverse wave components, respectively; $\theta_{ij}$ are angles of reflection ($j=1$) and angles of transmission ($j=2$); $k_{ij} = |\textbf k_{ij}| = \omega \hat n_{ij}/c$ are the wavenumbers and $\hat n_{ij} = n_j s_{ij}$ (no summation over $j$) are the effective refractive indices of these waves, which spatial dispersion is defined by the coefficients 
$s_{ij}= 1/\sqrt{1-l_{ij}^2\omega^2v^{-2}_j}$.

Taking into account gauge condition \eqref{gauge} in representation \eqref{helm}, \eqref{gs}, the electromagnetic potentials can be defined as follows:
\begin{equation}
\label{gsAo}
\begin{aligned}
z<0:\quad\textbf A_1 &= \Psi_0 \,\textbf e_{20}
+ \Psi_1 \,\textbf e_{21} - \tfrac{1}{\text i \omega}\nabla\tilde\phi_1,
\qquad \phi_1 = \tilde \phi_1 + \hat \phi_1\\
z>0:\quad\textbf A_2 &= \Psi_2 \,\textbf e_{22} - \tfrac{1}{\text i \omega}\nabla\tilde\phi_2,
\qquad \phi_2 = \tilde \phi_2 + \hat \phi_2 
\end{aligned}
\end{equation}
where all potentials are the functions of spatial coordinates $x$ and $z$; $\Psi_j$ ($j=0,1,2$), define the transverse components of incident, reflected and transmitted waves (we omit tilde symbol for these potentials), $\phi_j$ ($j=1,2$) define the longitudinal components of reflected and transmitted waves that contain the propagated $\tilde\phi_j$ and evanescent $\hat\phi_j$ parts; $\textbf e_{2j}$ ($j=0...2$) are the polarization unit vectors of transverse components of waves that lie in the $xz$-plane and that are perpendicular to wave vectors ($\textbf e_{2j}\cdot \textbf k_{2j}=0$). 
These unit vectors are defined as follows (see Fig. \ref{fig2}):
\begin{equation}
\label{eij}
\begin{aligned}
	\textbf e_{20} &= \cos \theta_{20} \,\textbf e_x - \sin \theta_{20} \,\textbf e_z\\
	\textbf e_{21} &= \cos \theta_{21} \,\textbf e_x + \sin \theta_{21} \,\textbf e_z\\
	\textbf e_{22} &= \cos \theta_{22} \,\textbf e_x - \sin \theta_{22} \,\textbf e_z
\end{aligned}
\end{equation} 

According to definition of general solution \eqref{gs}, potentials in \eqref{gsAo} should obey the following equations:
\begin{equation}
\label{gsAoeq}
\begin{aligned}
&(k_{2j} + \nabla^2) \Psi_j =0, \quad (j=0,1,2)\\
&(k_{1j} + \nabla^2) \tilde \phi_j =0, \quad (j=1,2)\\
&(1 - l^2_{1j}\nabla^2) \hat \phi_j =0, \quad (j=1,2)
\end{aligned}
\end{equation}

Taking into account definitions for wave vectors \eqref{k20}, \eqref{kij}, solutions for equations \eqref{gsAoeq} can be presented in the following form:
\begin{equation}
\label{poto}
\begin{aligned}
	\Psi_j (x,z) &= A_j \text e^{\text i \textbf k_{2j}\cdot \textbf x}, \quad (j=0,1,2)\\
	\tilde\phi_j(x,z) &= B_j \text e^{\text i \textbf k_{1j} \cdot \textbf x}, \quad (j=1,2)\\
	\hat\phi_j(x,z) &= C_j \text e^{\text i \textbf k_{3j} \cdot \textbf x}, \quad (j=1,2)
\end{aligned}
\end{equation}
where we formally introduce the wave vectors for evanescent waves $\textbf k_{3j}$ that will simplify us the satisfaction of boundary conditions. The corresponding wavenumbers are defined by $k_{3j} = |\textbf k_{3j}| = \text i/l_{1j}$ so that the solutions for $\hat\phi_j$ in \eqref{poto} will obey the  modified Helmholtz equations in \eqref{gsAoeq}. Similar to all other wave vectors, in the case of isotropic media, $\textbf k_{3j}$ should not contain projections onto the axis perpendicular to the plane of incidence (i.e. $(\textbf k_{3j})_y=0$).

From the solution of boundary value problem we should find six amplitudes $A_j$, $B_j$, $C_j$ ($j=1,2$) as well as the orientations of wave vectors $\textbf k_{ij}$ ($i=1,2,3$, $j=1,2$) that persist in \eqref{poto}. The amplitude $A_0$ and wave vector $\textbf k_{20}$ \eqref{k20} represent prescribed properties of the incident wave.

Due to the homogeneity of the solution along the $x$-axis, all potentials must depend on $x$ in the same manner. This yields the specular reflection law and generalized Snell's law for propagated waves:
\begin{equation}
\label{snel}
\begin{aligned}
	(\textbf k_{ij})_x &= (\textbf k_{20})_x
	\implies
	\theta_{21} = \theta_{20}, \quad 
	\frac{\sin \theta_{ij}}{\sin \theta_{20}} 
	= \frac{\hat n_{21}}{\hat n_{ij}},
	\quad (i,j =1,2)
\end{aligned}
\end{equation}
where we take into account definitions for wave vectors \eqref{k20}, \eqref{kij}.

For evanescent waves, based on the same condition, we directly define  the projections of wave vectors on $x$-axis:
\begin{equation}
\label{snele}
\begin{aligned}
	(\textbf k_{3j})_x 
	= (\textbf k_{20})_x
	= \frac{\omega}{c}\,\hat n_{21}\sin \theta_{20}
	\quad (j=1,2)
\end{aligned}
\end{equation}

Projections of wave vectors on $z$-axis in \eqref{kij} can be redefined in terms of prescribed angle $\theta_{20}$ by using \eqref{snel}, \eqref{snele}, as follows:
\begin{equation}
\label{kzo}
\begin{aligned}
	(\textbf k_{ij})_z &= \tfrac{\omega}{c}\sqrt{\hat n_{ij}^2 - \hat n_{21}^2 \sin^2\theta_{20}},
	\quad (i, j=1,2)
	\\
	(\textbf k_{3j})_z &= \text i\, (-1)^{j}\sqrt{\,l_{1j}^{-2} + \tfrac{\omega^2}{c^2}\hat n_{21}^2 \sin^2\theta_{20}},
	\quad (j=1,2)
\end{aligned}
\end{equation}
where for evanescent waves we take into account the correct direction of decay introducing factor $(-1)^j$.

From \eqref{kzo}, we have standard result for wave vector of reflected transverse wave: $(\textbf k_{21})_z = -(\textbf k_{20})_z$, $k_{21} = k_{20}$, while for evanescent wave we obtain purely imaginary $z$ projection for all frequencies that are possible within the dispersion relations of multipole theory \eqref{prop2}. 

By using \eqref{gsAo}, \eqref{poto} in relations \eqref{defE}, \eqref{defB}, we can obtain the following representations for electromagnetic fields in the first medium:
\begin{equation}
\label{gso01}
\begin{aligned}
	\textbf E_1&= 
	 \text i\omega\Psi_0 \,\textbf e_{20}
		+ \text i\omega \Psi_1 \,\textbf e_{21} 
		-2\text i\tilde\phi_1\textbf k_{11} 
		-\text i \hat\phi_1 \textbf k_{31}\\
	\textbf B_1 &= \mu_1 \textbf H_1
		= \text ik_{20}(\Psi_0+\Psi_1)\textbf e_y\\
\end{aligned}
\end{equation}
and in the second medium:
\begin{equation}
\label{gso02}
\begin{aligned}
	\textbf E_2&= 
	 \text i\omega \Psi_2 \,\textbf e_{22} 
		-2\text i\tilde\phi_2\textbf k_{12} 
		-\text i \hat\phi_2 \textbf k_{32}\\
	\textbf B_2& = \mu_2 \textbf H_2
		= \text ik_{22} \Psi_2 \textbf e_y\\
\end{aligned}
\end{equation}

The quadrupole tensor and electric displacement can be found then by using \eqref{ceq}, \eqref{D}.
The independent continuity conditions \eqref{bcc} at the boundary $z=0$ take the form:
\begin{equation}
\begin{aligned}
\label{BCo}
	[H_y + \text i\omega Q_{zx}]=0,\qquad
	[E_x] = 0,\qquad	
	[Q_{zz}] = 0,\\
	[D_z - \tfrac{\partial Q_{zx}}{\partial x}] = 0,\qquad
	[\phi] = 0,\qquad
	[\tfrac{\partial \phi}{\partial z}] = 0
\end{aligned}
\end{equation}
and the remaining continuity conditions related to the normal components of the magnetic induction and the tangential components of the vector potential are not independent. 

Note that we have obtained six continuity conditions \eqref{BCo} to determine the amplitude coefficients present in the sought solution \eqref{poto}, thus making the problem solvable.
If we neglect any component in the reflected and transmitted waves, the problem becomes unsolvable. For instance, if we completely disregard the presence of longitudinal components in the reflected and transmitted waves (see \cite{raab2005multipole} and references therein), the boundary conditions for the quadrupoles ($Q_{zz}$) and total electric displacement ($D_z - \partial Q_{zx}/\partial x$) will inevitably be violated, since even purely transverse waves lead to non-zero values of these field variables.
If we neglect the evanescent longitudinal waves but account for the propagating ones (like it was done in \cite{silveirinha2014boundary, yaghjian2016additional}), we introduce two additional amplitudes but will be unable to satisfy the continuity conditions for the scalar potential and its normal gradient, which become non-trivial.
Even when the first medium is a vacuum, we must account for four types of generated waves: the reflected transverse wave in the vacuum, and three types of waves in the second medium (transverse, propagating longitudinal, and evanescent longitudinal). This is necessary to satisfy the four non-trivial boundary conditions at the interface between vacuum and quadrupolar medium, which in this case reduce to:
\begin{equation}
\begin{aligned}
\label{BCov}
	[H_y + \text i\omega Q_{zx}]=0,\quad
	[E_x] = 0,\quad	
	Q_{zz} = 0,\quad
	D_z - \tfrac{\partial Q_{zx}}{\partial x} = 0
\end{aligned}
\end{equation}

Thus, a complete analytical solution for the problem under consideration had not been previously constructed accounting for the all necessary boundary conditions. While this solution can be found analytically based on the presented formulation \eqref{k20}-\eqref{BCo}, the resulting expressions for the amplitudes are rather cumbersome. Therefore, we will present only the results of numerical calculations for the oblique incidence problem. The calculations are performed as follows.
Substituting \eqref{gsAo}, \eqref{gso01}, \eqref{gso02} into \eqref{ceq},  \eqref{D}, and then into \eqref{BCo}, yields a system of equations for determining the unknown amplitudes $A_j$, $B_j$, $C_j$ ($j=1,2$). Using this solution in Eqs. \eqref{gso01}, \eqref{gso02}, and \eqref{ceq}, we compute the field variables required to evaluate the time-averaged Poynting vectors $\langle\textbf{S}_j\rangle$ ($j=0,1,2$) according to the general definition \eqref{Sg}. The reflectance and transmittance are then obtained from the normal components of these vectors as $R = (\langle\textbf{S}_1\rangle)_z/(\langle\textbf{S}_0\rangle)_z$ and $T = (\langle\textbf{S}_2\rangle)_z/(\langle\textbf{S}_0\rangle)_z$. The relative energy of evanescent wave is  formally assessed through the $x$-axis component of its Poynting vector as $(\langle\textbf{S}_3\rangle)_x/(\langle\textbf{S}_0\rangle)_z$.
 
\begin{figure}[h!]
 \centering
  (a)\includegraphics[width=0.4\linewidth]{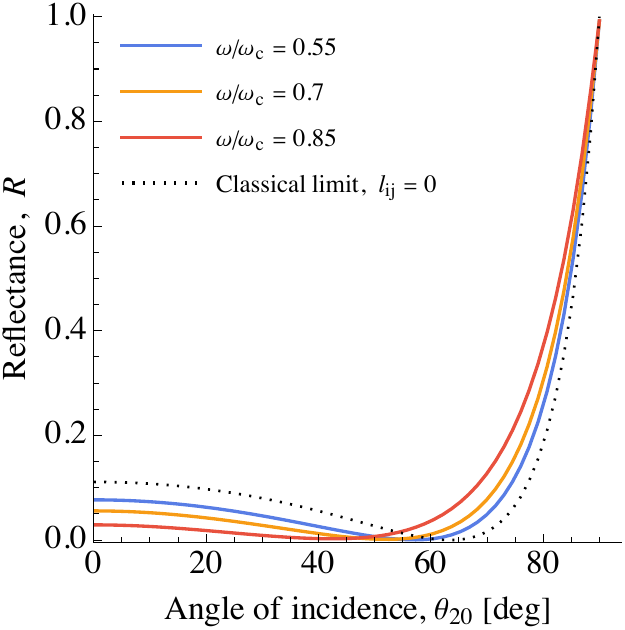}\quad
  (b)\includegraphics[width=0.4\linewidth]{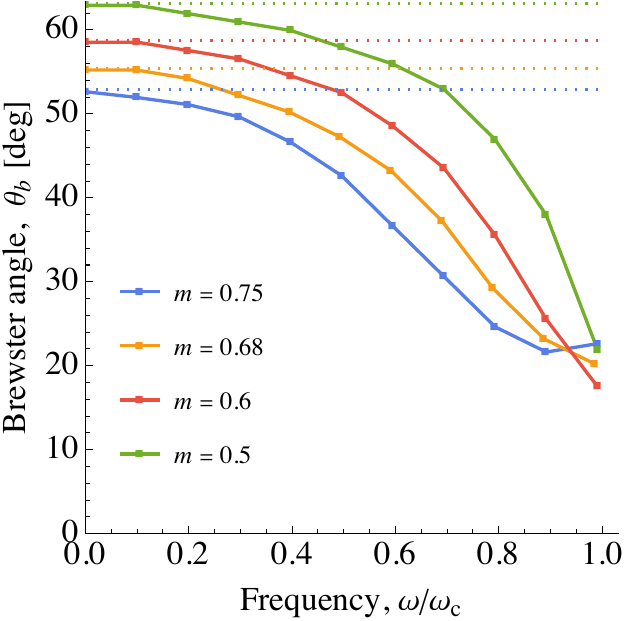}\\[10pt]
  (c)\includegraphics[width=0.4\linewidth]{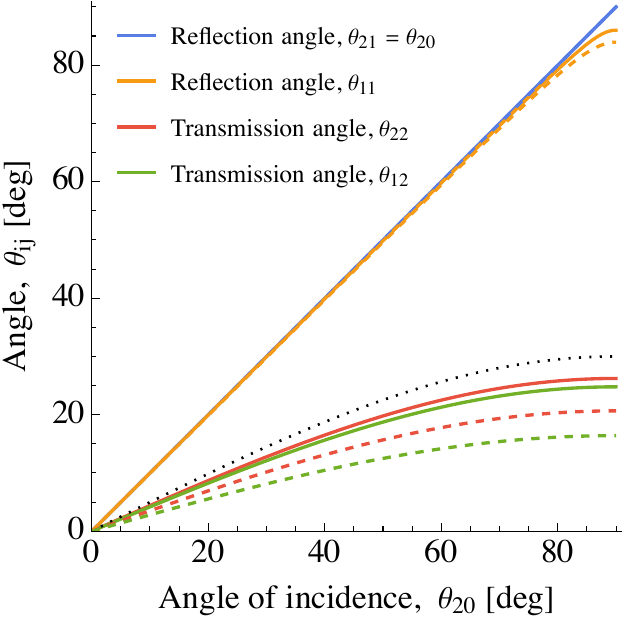}\quad
  (d)\includegraphics[width=0.4\linewidth]{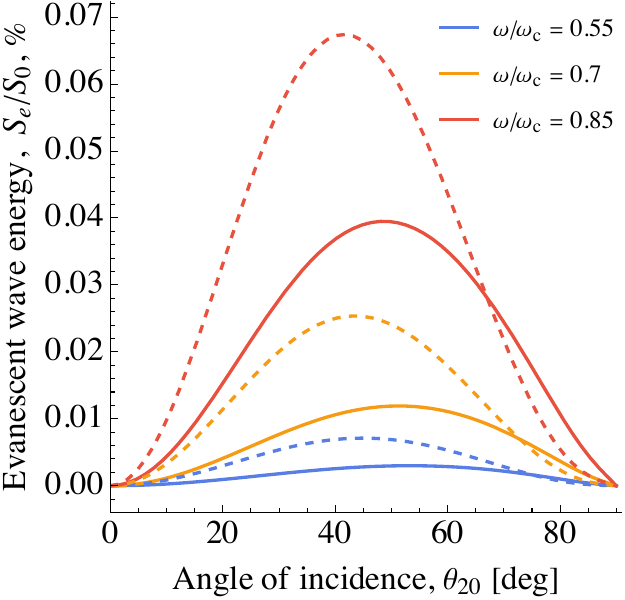}
  \caption{Solution of the oblique incidence problem. (a): Dependence of reflectance on indecent angle, (b): Dependence of Brewster angle on normalized frequency, (c): Dependence of reflection and transmission angles of transverse ($\theta_{21}$, $\theta_{22}$) and longitudinal ($\theta_{11}$, $\theta_{12}$) waves for normalized frequency value $\omega/\omega_c = 0.55$ (solid lines) and $\omega/\omega_c = 0.85$ (dashed lines), (d): Dependence of normalized surface energy of evanescent waves on incedence angle in medium 1 (solid lines) and in medium 2 (dashed lines). Classical solutions in absence of weak spatial dispersion are presented in all plots by dotted lines.}
  \label{fig4}
\end{figure}

The examples of calculations are presented in Fig. \ref{fig4}. These results are found for the case of traceless quadrupolar polarization tensor ($l_{1j} = 2l_{2j}/\sqrt{3}$), refractive indices of materials $n_1 = 1.5$ and $m=n_2/n_1=2$ and for the length scale parameters ratio $l_{22}/l_{21} = 2$.
Figure 4a shows the obtained dependence of the reflection coefficient on the angle of incidence. It can be seen that with increasing frequency (normalized to the cutoff frequency of the second layer $\omega_c = v_2/l_{12}$), $R$ increases at large incidence angles and decreases at small angles. Simultaneously, the Brewster angle (where $R=0$) shifts toward smaller incidence angles. The frequency dependence of the Brewster angle for different $m$ ratios is presented in Figure 4b. This dependence was determined through numerical analysis of results obtained from the multipole theory solution. The data show that as the frequency approaches the cutoff value, the Brewster angle in the quadrupole medium decreases to approximately 20-30 degrees, which appears to be a characteristic indicator for validating the proposed solutions when compared with experimental or computational data for metamaterials.

The propagation directions of transverse and longitudinal waves are shown in Fig. \ref{fig4}c according to the generalized Snell's law \eqref{snel}. Here, it can be observed that the propagation directions of longitudinal and transverse waves are quite close, particularly when operating away from the cutoff frequency. For reflected waves, this difference becomes significant only at angles exceeding 80 degrees. Furthermore, the propagation angles of longitudinal reflected and transverse waves are always smaller than the corresponding angles for transverse waves.
The relative energy of evanescent waves proves to be relatively small for the considered process parameters, not exceeding 0.1\% of the incident wave energy even for the high-frequency processes (Fig. 4d). The maximum surface wave energy occurs at incidence angles around 45 degrees and is higher in the medium with the larger length scale parameter. In this case, this is the second medium (see dashed lines in Fig. \ref{fig4}d).

%%%%%%%%%%%%%%%%%%%%%%%%%%%%%%%%%%%%%%%%%%%%%%
\section{Conclusions}

In this work, we developed a novel representation of the solution of Maxwell's equations  in terms of electromagnetic potentials within the framework of electrodynamics with multipoles. This solution \eqref{gs} satisfies uncoupled generalized wave equations that are derived using the suggested modified Lorentz gauge condition \eqref{ga} and operator form of constitutive equations \eqref{D}. Developed potential-based formalism enables convenient analytical computations for a broad class of multipole theory problems by reducing them to combinations of solutions to standard Helmholtz-type equations. Furthermore, we demonstrate that the proposed approach can be generalized to more general constitutive relations beyond the isotropic case considered. A clear examples illustrate the effective application of the developed approach.

We obtain a closed-form solution for the normal incidence problem that reliably validates correct approaches for calculating the Poynting vector in multipole theory. A noteworthy aspect of this solution is that the effective relative refractive index is not simply the ratio of the effective refractive indices of the two quadrupolar media, but exhibits a more complex dependence on their length scale parameters and hyper-susceptibilities \eqref{sysn}.

For oblique incidence, we show that complete solution of the problem requires consideration of the full set of boundary conditions: six conditions for the interface between two quadrupolar media or four conditions for the interface between a quadrupolar medium and vacuum. Such a complete formulation and solution have not been previously addressed. The from of the boundary conditions is derived through a variational approach that has also remained understudied in multipole theory. However, this approach enables reliable derivation of all essential and natural boundary conditions, ultimately determining the required continuity conditions for oblique incidence problem. 

We demonstrate that potential-based approach allows explicit representation of wave types arising from reflection/transmission phenomena in quadrupolar media. Specifically, it enables detailed analysis of longitudinal propagating and evanescent components of electromagnetic waves, providing precise predictions for reflection and transmission coefficients in quadrupolar materials. These results can be used for experimental validation of the theory.\\

\textbf{Acknowledgements}
This work was supported by the Russian Science Foundation grant number 23-11-00275.

%%%%%%%%%%%%%%%%%%%%%%%%%%%%%%%%%%%%%%%%%%%%%%
\appendix
\section*{Appendix A. Derivation of field equations and boundary conditions}

We consider the derivation of field equations and boundary conditions in multipole theory based on the least action principle. Note that this derivation is analogous to approaches used in other generalized continuum theories, where the internal energy density depends on higher-order spatial derivatives of the primary variables \cite{ eremeyev2018applications}. We consider the variation of the action functional represented by relation \eqref{S}:
\begin{equation}
\label{Sa}
\begin{aligned}
	&\delta \mathcal S =0, \qquad
	\mathcal S= 
	\int_{t_0}^{t_1}\int_\Omega \mathcal L dV dt,\\
	&\delta \mathcal S = \int_{t_0}^{t_1}\int_\Omega 
	\Big(
		-\varepsilon_0\textbf E\cdot\left(\delta\dot{\textbf A}+\nabla\delta\phi\right)
	- \tfrac{1}{\mu_0}\textbf B\cdot\left(\nabla\times\delta\textbf A\right)	
	-\rho\,\delta\phi + \textbf J\cdot\delta\textbf A\\
	&-\textbf P\cdot\left(\delta\dot{\textbf A}+\nabla\delta\phi\right) 
	+ \textbf M\cdot(\nabla\times\delta\textbf A)
	-\textbf Q:\left(\nabla\delta\dot{\textbf A}+\nabla\nabla\delta\phi\right)
	\Big)
	 dV dt
\end{aligned}
\end{equation}

We group the classical terms associated with polarization and magnetization:
\begin{equation}
\label{S1}
\begin{aligned}
	&\delta \mathcal S = \int_{t_0}^{t_1}\int_\Omega 
	\Big(
		-(\varepsilon_0\textbf E+\textbf P)\cdot\left(\delta\dot{\textbf A}+\nabla\delta\phi\right)
	- (\tfrac{1}{\mu_0}\textbf B-\textbf M)\cdot\left(\nabla\times\delta\textbf A\right)\\
%	\\-&\rho_f\delta\phi +  \textbf J_f\cdot\delta\textbf A 
	&-\textbf Q:\left(\nabla\delta\dot{\textbf A}+\nabla\nabla\delta\phi\right)
	-\rho\,\delta\phi + \textbf J\cdot\delta\textbf A
	\Big)
	 dV dt
\end{aligned}
\end{equation}

Then we use standard definition for magnetic field $\textbf H$ and apply the divergence theorem to quadrupole term $\textbf Q$ to obtain:
\begin{equation}
\label{S2}
\begin{aligned}
	&\delta \mathcal S = \int_{t_0}^{t_1}\int_\Omega 
	\Big(
		-(\varepsilon_0\textbf E+\textbf P-\nabla\cdot\textbf Q)\cdot\left(\delta\dot{\textbf A}+\nabla\delta\phi\right)	\\
		&\hspace{2.3cm}
		- \textbf H\cdot\left(\nabla\times\delta\textbf A\right)
		-\rho\,\delta\phi + \textbf J\cdot\delta\textbf A
	\Big)
	 dV dt \\
	 &+ \int_{t_0}^{t_1}\int_{\partial\Omega} 
	\Big(
		-\textbf n\cdot\textbf Q\cdot\left(\delta\dot{\textbf A}+\nabla\delta\phi\right)
	\Big)
	 dS dt
\end{aligned}
\end{equation}
where $\textbf n$ is outward unit vector on the body boundary $\partial \Omega$.

Subsequently, we introduce definition for the total electric displacement $\textbf D=\varepsilon_0\textbf E+\textbf P-\nabla\cdot\textbf Q$ and apply the divergence theorem to the magnetic field term:
\begin{equation}
\label{S3}
\begin{aligned}
	&\delta \mathcal S = \int_{t_0}^{t_1}\int_\Omega 
	\Big(
		-\textbf D\cdot\left(\delta\dot{\textbf A}+\nabla\delta\phi\right)	- (\nabla\times\textbf H)\cdot\delta\textbf A
		-\rho\,\delta\phi + \textbf J\cdot\delta\textbf A
	\Big)
	 dV dt \\
	 &+ \int_{t_0}^{t_1}\int_{\partial\Omega} 
	\Big(
		-\textbf n\cdot\textbf Q\cdot\left(\delta\dot{\textbf A}+\nabla\delta\phi\right)
		+(\textbf n\times\textbf H)\cdot\delta\textbf A
	\Big)
	 dS dt
\end{aligned}
\end{equation}
in which we used identities $\nabla\cdot(\textbf H\times\delta\textbf A) = 
 (\nabla\times\textbf H)\cdot\delta\textbf A
-(\nabla\times\delta\textbf A)\cdot\textbf H$ (in volime integral) and $\textbf n\cdot(\textbf H\times\delta\textbf A) = (\textbf n\times\textbf H)\cdot\delta\textbf A$ (in surface integral).

Then, we provide integration by parts with respect to time for all terms related to $\dot{\textbf A}$ and obtain:
\begin{equation}
\label{S4}
\begin{aligned}
	&\delta \mathcal S = \int_{t_0}^{t_1}\int_\Omega 
	\Big(
		-\textbf D\cdot\nabla\delta\phi	+ (\dot{\textbf D}-\nabla\times\textbf H)\cdot\delta\textbf A
		-\rho\,\delta\phi + \textbf J\cdot\delta\textbf A
	\Big)
	 dV dt \\
	 &+ \int_{t_0}^{t_1}\int_{\partial\Omega} 
	\Big(
		-\textbf n\cdot\textbf Q\cdot\nabla\delta\phi
		+\left(\textbf n\times\textbf H+\textbf n\cdot\dot{\textbf Q}\right)\cdot\delta\textbf A
	\Big)dS dt\\
	 &- \Big(\int_{\Omega} 
	\textbf D\cdot\delta\textbf A
	 dV\Big)_{t_0}^{t_1}
	 -\Big(\int_{\partial\Omega} 
		\textbf n\cdot\textbf Q\cdot\delta\textbf A 
	dS \Big)_{t_0}^{t_1}
\end{aligned}
\end{equation}
where the obtained time-boundary terms determine the form of initial conditions that should be defined within the initial-boundary value problem of multipole theory.

In volume integral in \eqref{S4} we apply the divergence theorem for term $\textbf D\cdot\nabla\delta\phi$ and obtain:
\begin{equation}
\label{S45}
\begin{aligned}
	&\delta \mathcal S = \int_{t_0}^{t_1}\int_\Omega 
	\Big(
		(\nabla\cdot\textbf D-\rho)\delta\phi	+ (\dot{\textbf D}-\nabla\times\textbf H + \textbf J)\cdot\delta\textbf A
		%-\rho\,\delta\phi + \textbf J\cdot\delta\textbf A
	\Big)
	 dV dt \\
	 &+ \int_{t_0}^{t_1}\int_{\partial\Omega} 
	\Big(
		-\textbf n\cdot\textbf D\,\delta\phi
		-\textbf n\cdot\textbf Q\cdot\textbf n\,\delta(\partial_n\phi)\\
		&\hspace{1.95cm}-\textbf n\cdot\textbf Q\cdot\nabla_S\delta\phi
		+\left(\textbf n\times\textbf H+\textbf n\cdot\dot{\textbf Q}\right)\cdot\delta\textbf A
	\Big)dS dt\\
	 &- \Big(\int_{\Omega} 
	\textbf D\cdot\delta\textbf A
	 dV\Big)_{t_0}^{t_1}
	 -\Big(\int_{\partial\Omega} 
		\textbf n\cdot\textbf Q\cdot\delta\textbf A 
	dS \Big)_{t_0}^{t_1}
\end{aligned}
\end{equation}

The volume integral in \eqref{S45} now contains the terms that define Gauss's and Ampere's laws, while the surface integral terms need separate consideration.
Namely, we proceed with further integration of the surface quadrupole terms appearing in the surface integral. This represents a standard approach in higher-order continuum theories that must consider the non-independence of tangential derivatives of the field variables along the body surface \cite{auffray2015analytical, eremeyev2018applications}. Utilizing the decomposition $\nabla\delta\phi = \nabla_S\delta\phi+\textbf n\partial_n\delta\phi$ (where $\nabla_S$ is surface gradient operator and $\partial_n=\nabla\cdot\textbf n$), we obtain:
\begin{equation}
\label{S5}
\begin{aligned}
	&\delta \mathcal S = \int_{t_0}^{t_1}\int_\Omega 
	\Big(
		(\nabla\cdot\textbf D-\rho)\delta\phi	+ (\dot{\textbf D}-\nabla\times\textbf H + \textbf J)\cdot\delta\textbf A
	\Big)
	 dV dt \\
	 &+ \int_{t_0}^{t_1}\int_{\partial\Omega} 
	\Big(
		-\textbf n\cdot\textbf D\,\delta\phi
		-\textbf n\cdot\textbf Q\cdot\textbf n\,\delta(\partial_n\phi)\\
		&\hspace{1.95cm}-\textbf n\cdot\textbf Q\cdot\nabla_S\delta\phi
		+\left(\textbf n\times\textbf H+\textbf n\cdot\dot{\textbf Q}\right)\cdot\delta\textbf A
	\Big)dS dt\\
	 &- \Big(\int_{\Omega} 
	\textbf D\cdot\delta\textbf A
	 dV\Big)_{t_0}^{t_1}
	 -\Big(\int_{\partial\Omega} 
		\textbf n\cdot\textbf Q\cdot\delta\textbf A 
	dS \Big)_{t_0}^{t_1}
\end{aligned}
\end{equation}

The emerging term $(\textbf n\cdot\textbf Q\cdot\textbf n)$ corresponds to additional boundary condition that has been extensively discussed in multipole theory, though previously it was derived through another methods \cite{silveirinha2014boundary, yaghjian2016additional, yaghjian2020comment}.  
For the term containing the surface gradient of quadrupolarization tensor in \eqref{S5}, we can apply the following surface divergence theorem (see, e.g. \cite{maugin2013continuum}):
\begin{equation}
\label{sdt}
\begin{aligned}
	&\int_{t_0}^{t_1}\int_{\partial\Omega} 
	(\textbf n\cdot\textbf Q\cdot\nabla_S\delta\phi) \,dS dt
	=\int_{t_0}^{t_1}\int_{\partial\Omega} 
	(\nabla_S\cdot(\textbf n\cdot\textbf Q\delta\phi)
		-\nabla_S\cdot(\textbf n\cdot\textbf Q)\delta\phi) \,dS dt\\
		&=\int_{t_0}^{t_1}\int_{\partial\Omega} 
	(-K(\textbf n\cdot\textbf Q\cdot \textbf n)\delta\phi
		-\nabla_S\cdot(\textbf n\cdot\textbf Q)\delta\phi) \,dS dt
		+\int_{t_0}^{t_1}\int_{\partial\partial\Omega} 
	 [\textbf n\cdot\textbf Q\cdot \pmb \nu] \delta\phi \,dL dt
\end{aligned}
\end{equation}
where $K = - \nabla\cdot\textbf n$ is twice the mean curvature of the boundary $\partial \Omega$;
		the brackets $[...]$ denote the jump of the enclosed quantities across the edge; $\pmb \nu$ is the co-normal vector that is tangent to surface $\partial\Omega$ and normal to edge $\partial\partial\Omega$. Note that if the body boundary $\partial\Omega$ is smooth and  does not contain edges $\partial\partial\Omega$, then the integral along these edges should be avoided in \eqref{sdt}.

Finally, using \eqref{sdt} in \eqref{S5}, we obtain:
\begin{equation}
\label{S6}
\begin{aligned}
	\delta \mathcal S &= \int_{t_0}^{t_1}\int_\Omega 
	\Big(
		(\nabla\cdot\textbf D-\rho)\delta\phi	+ (\dot{\textbf D}-\nabla\times\textbf H + \textbf J)\cdot\delta\textbf A
	\Big)
	 dV dt \\
	 &+ \int_{t_0}^{t_1}\int_{\partial\Omega} 
	\Big(
		-(\textbf n\cdot\textbf D-\nabla_S\cdot(\textbf n\cdot\textbf Q) -KQ_{nn})\,\delta\phi\\
		&\qquad\qquad\qquad-Q_{nn}\,\delta(\partial_n\phi)
		+\textbf n\times \left(\textbf H+(\textbf n\cdot\dot{\textbf Q})\times\textbf n\right)\cdot\delta\textbf A
	\Big)dS dt\\
	 &- \Big(\int_{\Omega} 
	\textbf D\cdot\delta\textbf A
	 dV\Big)_{t_0}^{t_1}
	 -\Big(\int_{\partial\Omega} 
		\textbf n\cdot\textbf Q\cdot\delta\textbf A 
	dS \Big)_{t_0}^{t_1}
	- \int_{t_0}^{t_1}\int_{\partial\partial\Omega} 
	 [\textbf n\cdot\textbf Q\cdot \pmb \nu] \delta\phi \,dL dt
\end{aligned}
\end{equation}
where we introduce notation $Q_{nn}=\textbf n\cdot\textbf Q\cdot \textbf n$ and 
take into account that this term should be zero on the free surface or it should continuous on the contact between two media, so that the following identity is valid:
$$
\textbf n\times\textbf H+\textbf n\cdot\dot{\textbf Q}
=\textbf n\times\textbf H
	+\textbf n\cdot\dot{\textbf Q}\cdot(\textbf I - \textbf n\otimes\textbf n)
= \textbf n\times \left(\textbf H+(\textbf n\cdot\dot{\textbf Q})\times\textbf n\right)\cdot\delta\textbf A
$$
in which the term $\textbf n\times (\textbf n\cdot\dot{\textbf Q})\times\textbf n$ is obtained based on the triple vector product rule:
$$\textbf n\times (\textbf n\cdot\dot{\textbf Q})\times\textbf n = 
(\textbf n\cdot\textbf n)(\textbf n\cdot\dot{\textbf Q}) - (\textbf n\cdot\dot{\textbf Q}\cdot\textbf n)\textbf n = 
\textbf n\cdot\dot{\textbf Q}\cdot(\textbf I - \textbf n\otimes\textbf n)$$

Thus, the obtained form of the variation of action \eqref{S6} is equivalent to those one used in the main text of the paper \eqref{Sder}, where we avoid the terms related to the initial conditions.

%%%%%%%%%%%%%%%%%%%%%%%%%%%%%%%%%%%%%%%%%%%%%%
\appendix
\section*{Appendix B. Definition of Poynting vector.}

In this Appendix we derive the Poynting theorem and definition of Poynting vector by using Maxwell equations of multipole theory \eqref{fe}. We start with identity $\nabla\cdot(\textbf E\times \textbf H)= 
	(\nabla\times\textbf E)\cdot \textbf H
	- (\nabla\times\textbf H)\cdot \textbf E$, in which we can use Faraday law \eqref{fea} and Amp\'ere law \eqref{fed} together with constitutive equations for $\textbf D$ field \eqref{D}. As the result, we obtain:
\begin{equation}
\label{b1}
\begin{aligned}
	\nabla\cdot(\textbf E\times \textbf H) &= 
	- \dot{\textbf B}\cdot\textbf H
	- \dot{\textbf D}\cdot\textbf E
	- \textbf J \cdot\textbf E \\
	&= -\mu_0\dot{\textbf H}\cdot \textbf H
	- \varepsilon_0\dot{\textbf E}\cdot\textbf E
	- \dot{\textbf P}\cdot\textbf E
	+ (\nabla\cdot\dot{\textbf Q})\cdot\textbf E
	- \textbf J \cdot\textbf E 
\end{aligned}
\end{equation}

By using standard vector identities, from \eqref{b1} we find:
\begin{equation}
\label{b2}
\begin{aligned}
	\nabla\cdot(\textbf E\times \textbf H) &= -\tfrac{1}{2}\tfrac{\partial}{\partial t}(\mu_0 \textbf H^2+ \varepsilon_0 \textbf E^2)
	- \dot{\textbf P}\cdot\textbf E
	+ \nabla\cdot(\dot{\textbf Q}\cdot\textbf E)
	- \dot{\textbf Q}:\nabla\textbf E
	- \textbf J \cdot\textbf E 
\end{aligned}
\end{equation}
and therefore:
\begin{equation}
\label{b3}
\begin{aligned}
	\nabla\cdot(\textbf E\times \textbf H - \dot{\textbf Q}\cdot\textbf E) 
	&= -\tfrac{1}{2}\tfrac{\partial}{\partial t}
		(\mu_0 \textbf H^2+ \varepsilon_0 \textbf E^2)
	- \dot{\textbf P}\cdot\textbf E
	- \dot{\textbf Q}:\nabla\textbf E
	- \textbf J \cdot\textbf E 
\end{aligned}
\end{equation}

The obtained relation \eqref{b3} is exactly the generalization of Poynting theorem for multipole theory, that can be rewritten as follows:
\begin{equation}
\label{b23}
\begin{aligned}
	\tfrac{\partial u}{\partial t} + \nabla\cdot\textbf S + \textbf J \cdot\textbf E= 0
\end{aligned}
\end{equation}
where $u = \tfrac{1}{2}(\mu_0 \textbf H^2+ \varepsilon_0 \textbf E^2)
	+ \tfrac{1}{2}{\textbf P}\cdot\textbf E
	+ \tfrac{1}{2}{\textbf Q}:\nabla\textbf E$ is the energy density that contains additional term related to quadrupole polarization (see \cite{slavchov2014quadrupole}) and the generalized definition for Poynting vector is given by: 
\begin{equation}
\label{b4}
\begin{aligned}
	\textbf S = \textbf E\times \textbf H - \dot{\textbf Q}\cdot\textbf E
\end{aligned}
\end{equation}

This definition \eqref{b4} coincides with those one derived in Refs. \cite{yaghjian2016additional,yaghjian2018power} based on the spatial averaging procedure for energy flux accounting for spatial dispersion effects. The corresponding time-average quantity in time-harmonic processes is given by \cite{yaghjian2016additional,yaghjian2018power}: 
\begin{equation}
\label{b5}
\begin{aligned}
	\langle\textbf S\rangle &=  
	\frac{1}{2}\text{Re}(\textbf E\times\textbf H^* 
		+ \text i \omega \,\textbf E^*\cdot\textbf Q), \\
\end{aligned}
\end{equation}

Considering the propagation phenomena in infinite medium in Section 4 we used the absolute value   $S=|\langle\textbf S\rangle|$. Substituting \eqref{propf}-\eqref{imp} into \eqref{b5} it can be defined as:
\begin{equation}
\label{b6}
\begin{aligned} 
	S &= \frac{1}{2}
	\left(
		\frac{E^2_\perp s_2}{Z} 
		- \omega k_2 l_2^2\varepsilon E^2_\perp
		- \omega k_1 l_1^2\varepsilon E^2_\|
			\right)
\end{aligned}
\end{equation} 
where $Z = \sqrt{\mu/\varepsilon}$ and the last two terms corresponds to the influence of quadrupole effects related to the transverse and longitudinal components of electric field.

By using dispersion relations \eqref{prop2} and definitions for effective refractive indices \eqref{hn} in \eqref{b6}, we finally obtain:
\begin{equation}
\label{b7}
\begin{aligned} 
	S &= 
	\frac{E^2_\perp}{2}\left(
		\frac{s_2}{Z} 
		- \frac{c}{\hat n_2} k^2_2 l_2^2\varepsilon
			\right)
		-\frac{E^2_\|}{2}\frac{c}{\hat n_1} k^2_1 l_1^2\varepsilon\\
	&= 
	\frac{E^2_\perp}{2}\left(
		\frac{s_2}{Z}  
		- \frac{1}{s_2} k^2_2 \sqrt{\frac{\varepsilon}{\mu}} l_2^2
			\right)
		- \frac{E^2_\|}{2}
		\frac{1}{s_1} k^2_1 \sqrt{\frac{\varepsilon}{\mu}} l_1^2\\
	&= 
	\frac{E^2_\perp}{2}\left(
		\frac{s_2}{Z}  
		- \frac{1}{Z s_2} k^2_2 l_2^2
			\right)
		- \frac{E^2_\|}{2Z s_1}k^2_1 l_1^2\\
	&= 
	\frac{E^2_\perp}{2Zs_2}\left(
		s_2^2 
		- k^2_2 l_2^2
			\right) -\frac{E^2_\|}{2 Zs_1}k^2_1 l_1^2\\
	&= 
	\frac{E^2_\perp}{2Zs_2} -\frac{E^2_\|}{2 Zs_1}k^2_1 l_1^2
\end{aligned}
\end{equation} 
where we also take into account the definition for coefficient $s_2 = \sqrt{1+l_2^2k_2^2}$.

\section*{References}
\renewcommand{\bibsection}{}
\bibliography{refs.bib}

\end{document}